\documentclass[fleqn,10pt]{wlscirep}
\usepackage[utf8]{inputenc}
\usepackage[T1]{fontenc}

\usepackage{graphicx}
\usepackage{dcolumn}
\usepackage{bm}
\usepackage{siunitx}
\usepackage{gensymb}

\usepackage{amsmath} 

\usepackage[caption=false]{subfig}
\usepackage{xcolor}
\usepackage{hyperref}
\usepackage{hyperref}
\hypersetup{
    colorlinks=true,
    linkcolor=blue,
    filecolor=blue,      
    urlcolor=blue,
    citecolor=blue,
}
\definecolor{gray}{rgb}{0.5,0.5,0.5}
\newcommand{\ulg}{\textcolor{gray}{$\blacktriangleright$}}

\DeclareUnicodeCharacter{2212}{-}

\title{Thin film epitaxial [111] Co$_{50}$Pt$_{50}$: Structure, magnetisation, and spin polarisation}

\author[1]{N.~Satchell}
\author[1]{S.~Gupta}
\author[1]{M.~Maheshwari}
\author[1]{P. M.~Shepley}
\author[1]{M.~Rogers}
\author[1]{O.~Cespedes}
\author[1,*]{G.~Burnell}
\affil[1]{School of Physics and Astronomy, University of Leeds, Leeds, LS2 9JT, United Kingdom}
\affil[*]{g.burnell@leeds.ac.uk}

\begin{abstract}

Ferromagnetic films with perpendicular magnetic anisotropy are of interest in spintronics and superconducting spintronics. Perpendicular magnetic anisotropy can be achieved in thin ferromagnetic multilayer structures, when the anisotropy is driven by carefully engineered interfaces. Devices with multiple interfaces are disadvantageous for our application in superconducting spintronics, where the current perpendicular to plane is affected by the interfaces. Robust intrinsic PMA can be achieved in certain Co$_x$Pt$_{100-x}$ alloys and compounds at any thickness, without increasing the number of interfaces. Here, we grow equiatomic Co$_{50}$Pt$_{50}$ and report a comprehensive study on the structural, magnetic, and spin-polarisation properties in the $L1_1$ and $L1_0$ ordered compounds. Primarily, interest in Co$_{50}$Pt$_{50}$ has been in the $L1_0$ crystal structure, where layers of Pt and Co are stacked alternately in the [100] direction. There has been less work on $L1_1$ crystal structure, where the stacking is in the [111] direction. For the latter $L1_1$ crystal structure, we find magnetic anisotropy perpendicular to the film plane. For the former $L1_0$ crystal structure, the magnetic anisotropy is perpendicular to the [100] plane, which is neither in-plane or out-of-plane in our samples. We obtain a value for the ballistic spin polarisation of the $L1_1$ and $L1_0$ Co$_{50}$Pt$_{50}$ to be $47\pm3\%$.

\end{abstract}
\begin{document}

\flushbottom
\maketitle
%
%
\thispagestyle{empty}


\section*{Introduction}

Ferromagnetic films with perpendicular magnetic anisotropy (PMA) are of wide interest for applications in established and nascent technologies such as ultrahigh density magnetic hard drives \cite{doi:10.1063/1.2750414}, MRAM\cite{wong2015memory}, superconducting spintronics \cite{linder2015superconducting}, and energy efficient spin-orbit torque memory \cite{RevModPhys.91.035004}. PMA can be achieved in Pt/Co/Pt multilayer systems as a result of the interfacial anisotropy, however at a critical Co thickness, typically about 1~nm, the anisotropy will fall in-plane. Increasing the total ferromagnetic layer thickness further therefore involves adding additional interfaces to the multilayer. Having a multilayer structure introduces interfacial resistance and interfacial spin-flip scattering \cite{bass2007spin}, which are disadvantageous for applications such as ours which require the transport current perpendicular-to-plane  \cite{birge2018spin, doi:10.1063/1.5140095 ,satchell2021pt}. Alternately, robust intrinsic PMA can be achieved in certain Co$_x$Pt$_{100-x}$ alloys and compounds at any thickness, without increasing the number of interfaces.   

Here, we study the equiatomic Co$_{50}$Pt$_{50}$ alloy, hereafter referred to as CoPt. Through growth at elevated temperatures it is possible to form the $L1_1$ and $L1_0$ chemically ordered compounds of CoPt as epitaxial films. Previous experimental studies of such compounds tend to use MgO substrates as the basis for high temperature epitaxial growth. Visokay and Sinclair report $L1_0$ crystal structure on MgO [001] substrates for growth temperatures above 520$^{\degree}$C \cite{doi:10.1063/1.113895}. Iwata \textit{et al.} report growth of $L1_1$ crystal structure on MgO [111] substrates for a growth temperature of 300$^{\degree}$C \cite{619533}.

Early thin film studies of the chemically ordered CoPt (and the related FePt and FePd) compounds were motivated by the large out-of-plane anisotropy and narrow domain wall widths being candidate for high density storage mediums \cite{619533, PhysRevB.50.3419, PhysRevB.52.13419, doi:10.1063/1.362122, doi:10.1063/1.361368, doi:10.1063/1.113895, doi:10.1063/1.364504, caro1998structure, doi:10.1063/1.368158, doi:10.1063/1.368831, doi:10.1063/1.371397, doi:10.1126/science.287.5460.1989, zeng2002orientation, PhysRevB.66.024413, chen2003effect, zhao2003promotion, doi:10.1021/jp027314o, LAUGHLIN2005383,  doi:10.1063/1.1991968, LIAO2006e243, doi:10.1063/1.2176306, PhysRevB.76.174435, doi:10.1063/1.2730568, Sato2008, chen2008review, 4492859, doi:10.1063/1.3153513, Seemann_2010, yuan2010effect, Antje_Dannenberg_2010, doi:10.1063/1.3561115,  doi:10.1063/1.3672856, ohtake2012structure, sun2013formation, 10.1063/1.4799526, 6416998, ohtake2014structure, VARVARO2014415, doi:10.1063/1.4960554, zygridou2017exploring, PhysRevApplied.10.054015, GAO2019406, doi:10.1021/acsanm.9b01192, PhysRevMaterials.6.024403, spencer2022characterization}. Recently, renewed interest in these compounds has been driven by the discovery of self-induced spin-orbit torque switching in these materials, which can be used as the switching mechanism for a low-dissipation magnetic memory \cite{PhysRevB.101.220402,https://doi.org/10.1002/adfm.202005201,doi:10.1063/5.0028815,liu2021symmetry,9433557,DONG2022,doi:10.1063/5.0077465, liu2022current}.

Our motivation for studying [111] CoPt is to incorporate this PMA ferromagnet in an all epitaxial heterostructure suitable for superconducting Josephson devices\cite{birge2018spin, doi:10.1063/1.5140095 ,satchell2021pt} or MRAM\cite{wong2015memory}. For these applications growth in the [111] direction is favourable. In Josephson junctions, the superconductor of choice is Nb, which can be grown epitaxially as a bcc structure in the [110] direction\cite{WILDES20017}. In MRAM, the seed layer of choice is Ta, which has almost identical structural properties to Nb. On Ta or Nb [110] layers, we know that Pt and Co will grow with [111] orientation\cite{PhysRevB.97.214509}. We therefore choose Al$_2$O$_3$ [0001] substrates for this study and use thin Pt [111] seed layers to grow CoPt [111]\cite{yuan2010effect}. As far as we are aware, no previous works have reported the properties of [111] ordered films of $L1_0$ and $L1_1$ CoPt on Al$_2$O$_3$ [0001] substrates. We are also unaware of any measurements of the spin polarisation of CoPt. 


We fabricate and report the properties of three sets of samples. The first set is designed to determine the optimal growth temperature. Therefore, we fix the thickness $d_{CoPt} = 40$~nm and vary the substrate heater temperature in the range from 27${\degree}$C to 850${\degree}$C. The next two sample sets are thickness series grown by varying the growth time at a fixed temperature and magnetron powers. The temperatures chosen for the thickness series are to produce either the $L1_1$ (350${\degree}$C) or $L1_0$ (800${\degree}$C) crystal structures. These temperatures are guided by the results of the first temperature series of samples. Thicknesses are varied over the range 1~nm$ \leq d_\text{CoPt} \leq 128$~nm.

On each sample set we report systematically on the structural and magnetic properties of the CoPt. Additionally, on the thickest 128~nm $L1_1$ and $L1_0$ samples we perform point contact Andreev reflection (PCAR) measurements with a Nb tip at 4.2~K to determine the spin polarisation. The use of Nb as the tip, the temperature of this measurement, and the ballistic transport regime probed are relevant for our proposed application of the CoPt in Josephson devices\cite{birge2018spin, doi:10.1063/1.5140095 ,satchell2021pt}.


\section*{Results and Discussion}

\subsection*{\label{Growth}CoPt properties as a function of growth temperature}

 We expect that as the growth temperature is increased the CoPt will form initially a chemically disordered $A1$ alloy phase, followed by a chemically ordered $L1_1$ crystal structure, an intermediate temperature $A1$ phase, and finally a chemically ordered $L1_0$ crystal structure respectively. In order to map out the growth parameters we report on Al$_2$O$_3$(sub)/Pt(4~nm)/CoPt(40~nm)/Pt(4~nm) sheet films grown at different set temperatures in the range from room temperature (RT) to 850$^{\degree}$C. 
 
 
\subsubsection*{Structure}

\begin{figure}[ht]
    \centering
    \includegraphics[width=0.9\textwidth]{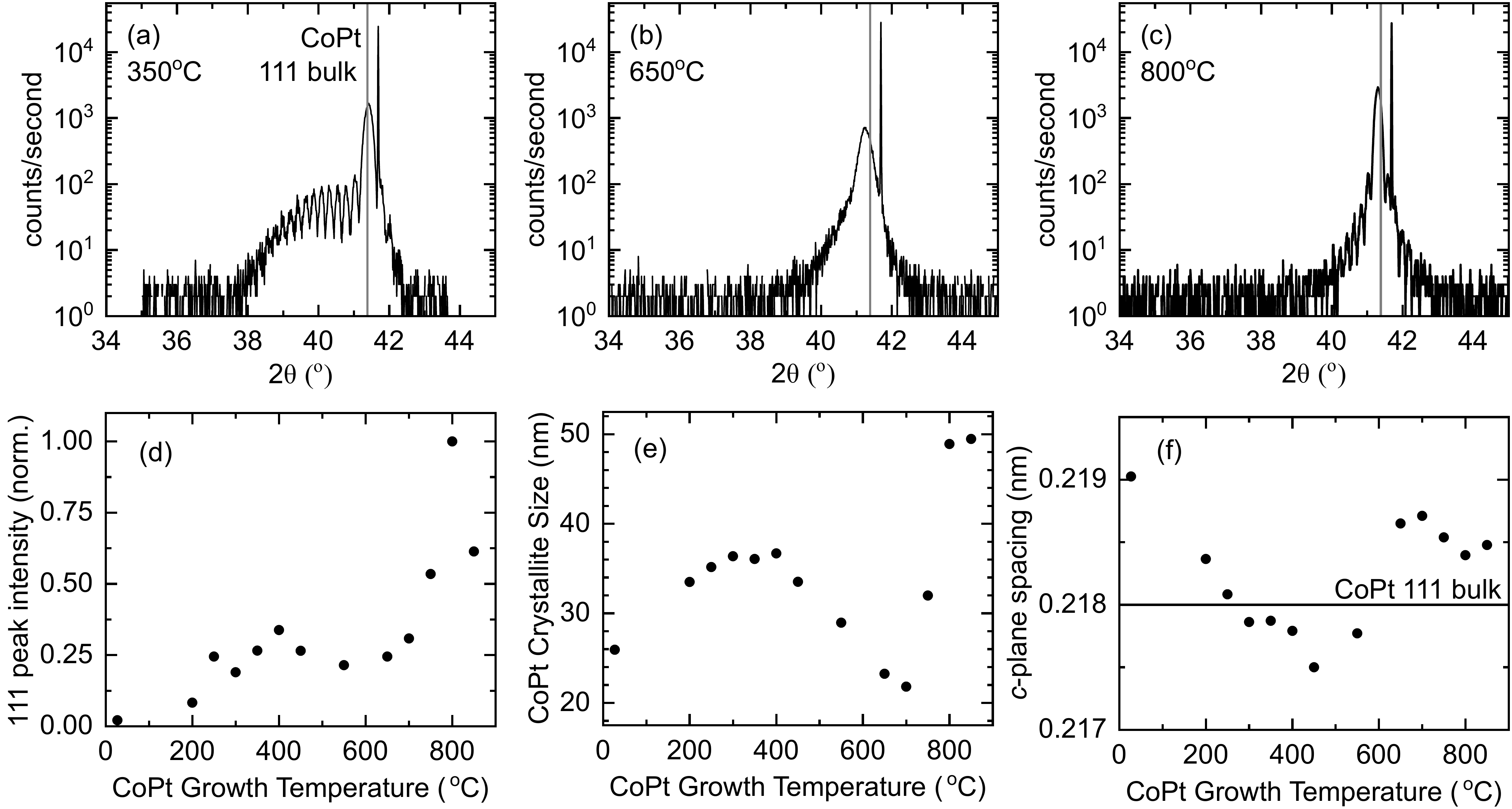}
    \caption{Structural characterisation of Al$_2$O$_3$(sub)/Pt(4~nm)/CoPt(40~nm)/Pt(4~nm) sheet films grown at different set temperatures obtained by x-ray diffraction measurements. (a-c) X-ray diffraction data for samples at selected growth temperatures. The position of the main [111] structural peak for the CoPt layer is indicated. (d) The normalised CoPt [111] peak intensity. (e) The CoPt crystallite size determined from the CoPt [111] peak Gaussian full width half maximum value. (f) The CoPt $c$-plane spacing. Uncertainties in the peak intensity, crystallite size, and $c$-plane spacing are smaller than the data symbols.}
    \label{fig:XRD}
\end{figure}

In order to understand the magnetic phases of sputter deposited CoPt, it is necessary to understand the underpinning structure and the influence of the growth temperature. Therefore, the structural characteristics and film quality were investigated using x-ray diffraction (XRD) as a function of growth temperature. Fig.~\ref{fig:XRD} (a-c) shows the acquired XRD for selected growth temperatures around the main CoPt [111] structural peak. Further XRD data are available in the Supplemental Information online. The CoPt [111] structural peak is clearly observable for all growth temperatures. The presented data at growth temperatures of 350$^{\degree}$C and 800$^{\degree}$C show additional Pendellosung fringes which bracket the primary CoPt [111] peak. The CoPt [111] relative peak intensity is shown as a function of growth temperature in Fig.~\ref{fig:XRD} (d). In Fig.~\ref{fig:XRD} (a), for a growth temperature of 350$^{\degree}$C, the additional feature at $2\theta\approx40^{\degree}$ corresponds to the superimposed Pendellosung fringes and the Pt [111] structural peak (bulk Pt has a lattice constant of 3.92\AA). This additional Pt structural peak is present for growths up to 550$^{\degree}$C (see Supplementary Information online) which is within the expected optimal temperature range for sputter deposited Pt films on Al$_2$O$_3$ \cite{4856395}. To maintain consistent Pt structure for CoPt growths at different temperatures, future work could grow the Pt layer at its optimum growth temperature.

Using the Scherrer equation and the full widths at half maxima of the Gaussian fits to the CoPt [111] structural peaks, an estimation of the CoPt crystallite sizes can be made. The crystallite size determined by the main [111] structural peak is given in Fig.~\ref{fig:XRD} (e). In the expected range of the ordered $L1_1$ crystal structure between 200$^{\degree}$C and 400$^{\degree}$C, the estimated CoPt crystallite size is 37~nm, which compared to the nominal thickness of 40~nm indicates that the CoPt has high crystallinity. On the other hand, at RT growth and intermediate temperatures between 400$^{\degree}$C and 800$^{\degree}$C, the estimated CoPt crystallite size is much lower, showing a minimum value of 22~nm at 700$^{\degree}$C where we expect $A1$ growth. Interestingly, the disorder in the $A1$ growth appears to affect both the chemical disorder (random positions of the Co and Pt atoms in the unit cell) and leads to a poorer crystallite size compared to the chemically ordered crystal structures. The disappearance of the Pendellosung fringes at these intermediate growth temperatures is related to the increased roughness of the $A1$ films. Finally, upon increasing the temperature further to 800$^{\degree}$C and 850$^{\degree}$C the crystallite size reaches a maximum of 49~nm. This value corresponds approximately to the entire thickness of the Pt/CoPt/Pt trilayer, indicating that the Pt buffer and capping layers have fully interdiffused into the CoPt layer. 

Fig.~\ref{fig:XRD} (f) shows the CoPt $c$-plane space calculated from the center of the fitter Gaussian to the main [111] structural peak. At all temperatures the measured $c$-plane spacing is very close to the expected value based on single crystal studies \cite{mccurrie1969magnetic}, indicating low out-of-plane strain. The trend with temperature, however, is non-monotonic and shows a discrete increase between the samples grown at 550$^{\degree}$C and 650$^{\degree}$C. The transition associated with the $L1_0$ crystal structure is expected to result in a transition from cubic to tetragonal crystal with a $c/a=0.979$ in the single crystal \cite{mccurrie1969magnetic}. In the single crystal study, however, the $c$-axis is in the [001] orientation. In our [111] orientated film, the small structural transition from cubic to tetragonal crystal structure is not expected to be visible in the [111] peak studied here. Nonetheless, Fig.~\ref{fig:XRD} (f) shows features consistent with the CoPt undergoing structural transitions.

\begin{figure}[ht]
    \centering
    \includegraphics[width=0.7\textwidth]{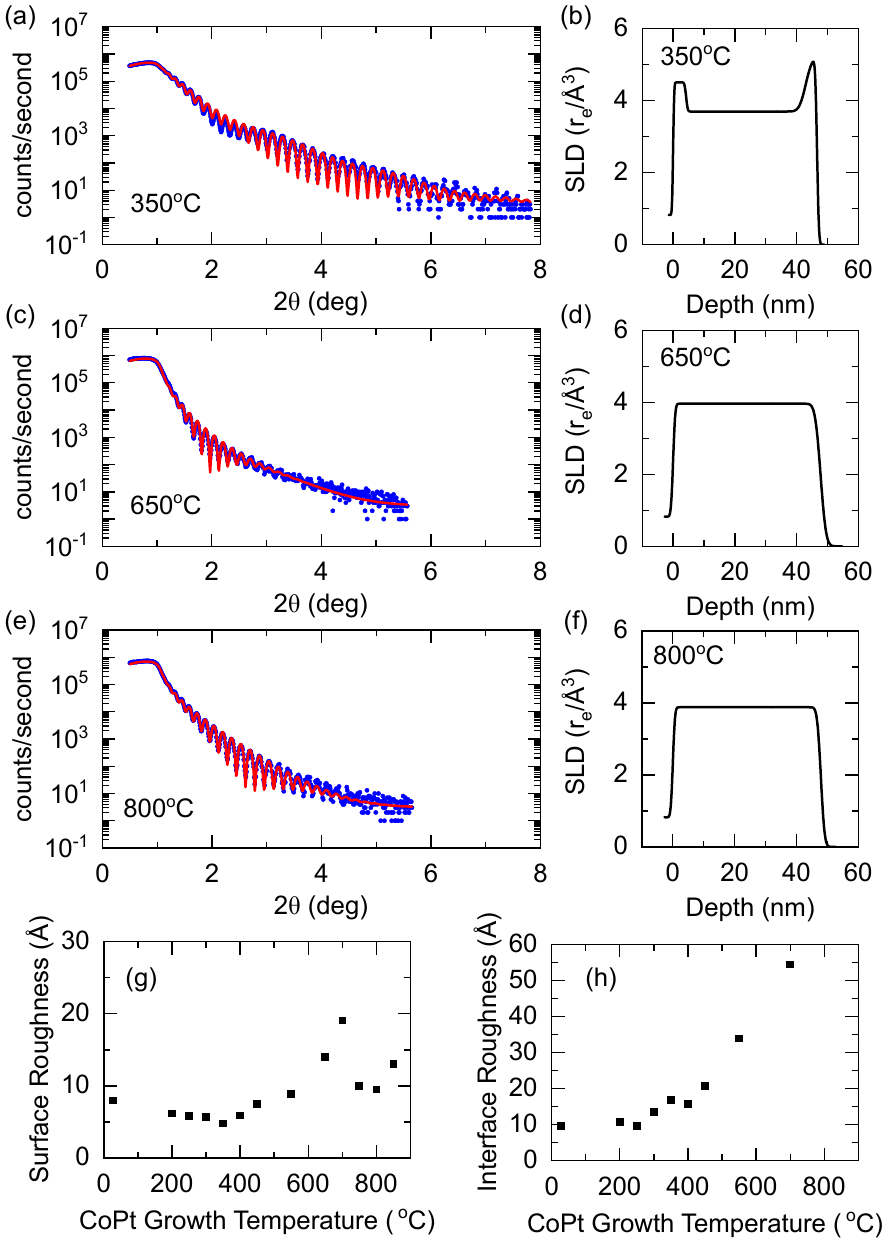}
    \caption{Structural characterisation of Al$_2$O$_3$(sub)/Pt(4~nm)/CoPt(40~nm)/Pt(4~nm) sheet films grown at different set temperatures obtained by x-ray reflectivity measurements. (a,c,e) X-ray reflecitvity data for samples at selected growth temperatures. The reflectivity is fit to a best fit model, which is shown correspondingly in (b,d,f). (g) The surface roughness parameter extracted from the best fit model. (h) The interfacial roughness parameter from the model, indicating the interdiffusion between the CoPt and Pt layers.}
    \label{fig:XRR}
\end{figure}

X-ray reflectivity (XRR) is performed to further investigate the growth temperature dependent trends in the structural properties of the films. Fig.~\ref{fig:XRR} (a,c,e) show the low angle XRR for selected growth temperatures along with the best fits to the data. The corresponding models are shown in Fig.~\ref{fig:XRR} (b,d,f). Further XRR data are available in the Supplemental Information online. The fitting is performed using the GenX package \cite{Bjorck:aj5091}, which models each layer as a box with independent thickness, roughness, and density fitting parameters. Across 13 samples grown at varying temperatures, the average total sample thickness is $47.2\pm0.3$ nm, compared to a nominal total thickness of 48.0 nm, confirming the sputter rate calibration and that the growth temperature has not impacted the growth rate. For growth temperatures of 450$^{\degree}$C and below, the XRR is best modelled as a Pt/CoPt/Pt trilayer, for example Fig. ~\ref{fig:XRR} (a,b). At a growth temperature of 550$^{\degree}$C and above there is a clear change in the XRR. Modelling the data suggests that the interdiffusion has become so large that a trilayer model is no longer required. For the data shown in Fig. ~\ref{fig:XRR} (c-f), a single layer model is used to fit the reflectivity.

Fig.~\ref{fig:XRR} (g) shows the extracted roughness parameter on the top surface of the sample. The surface roughness shows temperature dependence corresponding to the underpinning structure of the CoPt film. For the optimal growth temperature to achieve the $L1_1$ crystal structure, the corresponding surface roughness is lowest. In the $A1$ growth regime, the disorder in the crystal structure evident in the XRD is also present in the film surface roughness. The lower roughness is recovered at higher temperatures at the optimal growth temperature for the $L1_0$ crystal structure. Fig.~\ref{fig:XRR} (h) shows an extracted interfacial roughness parameter at the CoPt/Pt interface. We interpret this parameter as a measure of the interdiffusion between the layers. As expected, a clear trend is present where the interdiffusion between the layers increases with increasing temperatures. At the highest growth temperatures where a single layer model is used to fit the data, there is no interfacial roughness parameter to extract as the layer has completely interdiffused.


\subsubsection*{Magnetic characterisation}

\begin{figure}
    \centering
    \includegraphics[width=0.9\textwidth]{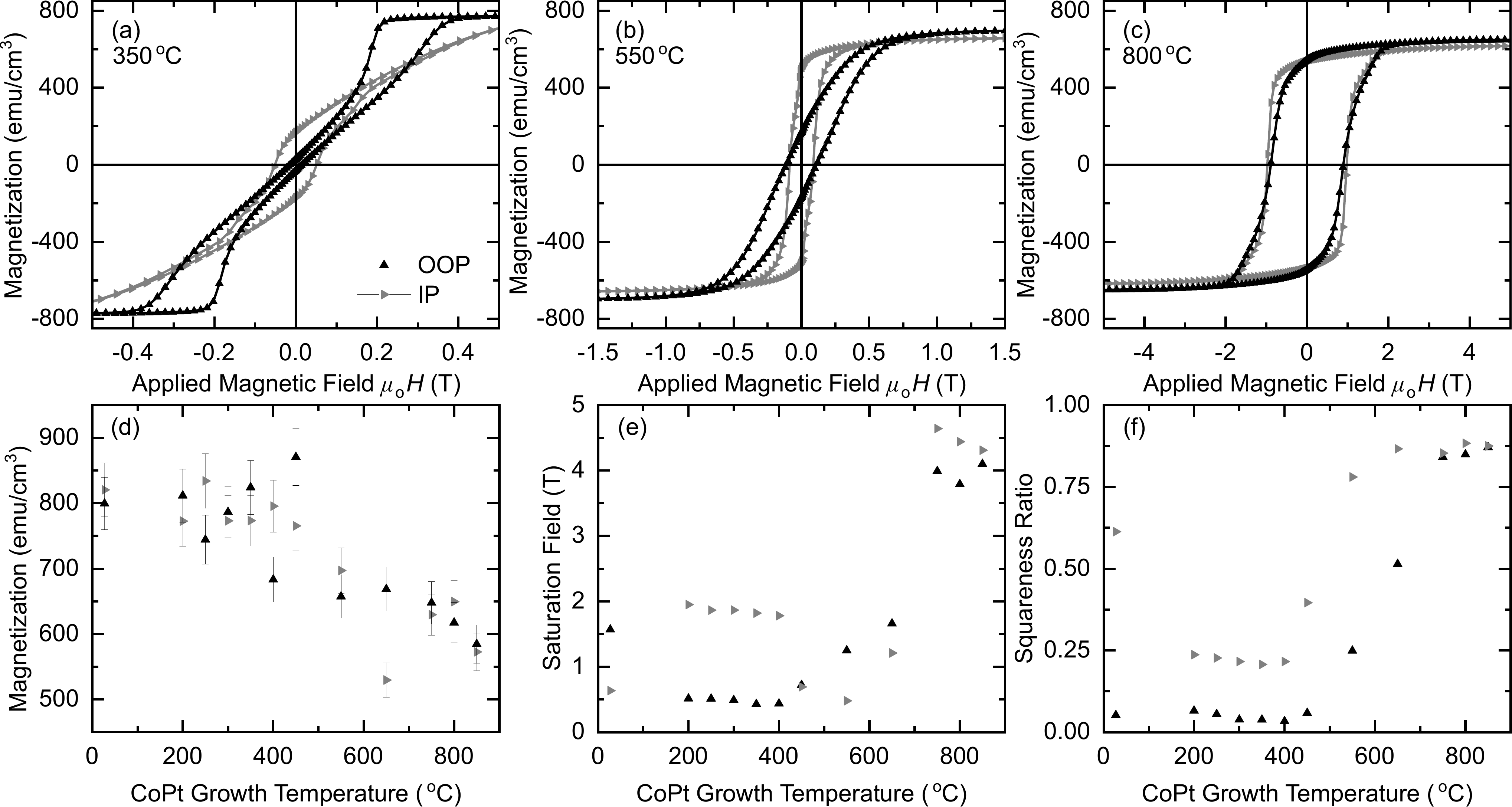}
    \caption{Magnetisation characterisation of Al$_2$O$_3$(sub)/Pt(4~nm)/CoPt(40~nm)/Pt(4~nm) sheet films grown at different set temperatures with the applied field oriented out-of-plane ($\blacktriangle$) and in-plane (\ulg). (a-c) Magnetic hysteresis loops acquired at a temperature of 300~K. The sample growth temperature is (a) 350${\degree}$C, corresponding to $L1_1$ crystal structure, (b) 550${\degree}$C, corresponding to disordered $A_1$ growth, and (c) 800${\degree}$C, corresponding to $L1_0$ crystal structure. The diamagnetic contribution from the substrate has been subtracted. (d-f) Trends in the magnetic characterisation as a function of sample growth temperature, (d) magnetisation, (e) saturation field and (f) squareness ratio ($M_r/M_s$). Magnetisation is calculated from the measured total magnetic moments, the areas of the sample portions, and the nominal thickness of the CoPt layer (40~nm). The significant contribution to the uncertainty in magnetisation is from the area measurements of the sample portions.}
    \label{fig:Magnetization_Growth_T}
\end{figure}

The magnetisation versus field data are shown in Fig. \ref{fig:Magnetization_Growth_T} for Al$_2$O$_3$(sub)/Pt(4 nm)/CoPt(40 nm)/Pt(4 nm) sheet film samples. Further magnetisation data for all samples in this study are available in the Supplemental Information online. The 350${\degree}$C, 550${\degree}$C, and 800${\degree}$C samples are plotted here as they are representative of the magnetic response of the $L1_1$ crystal structure, the chemically disordered $A1$ phase, and $L1_0$ crystal structure respectively. Magnetisation is calculated from the measured total magnetic moments, the areas of the sample portions, and the nominal thicknesses of the CoPt layer. 

For the chemically ordered $L1_1$ crystal structure shown in Fig. \ref{fig:Magnetization_Growth_T} (a), the OOP hysteresis loop shows a wasp waisted behavior associated with the formation of magnetic domains at remanence. Such behavior is common in CoPt alloys and multilayer thin films \cite{doi:10.1063/1.113895}. The wasp waisted OOP hysteresis loop along with the low IP remanence and higher IP saturation field indicates that the 40~nm CoPt samples with the $L1_1$ crystal structure have strong PMA. We can estimate the effective anisotropy using the expression $K_\text{eff} = \mu_0M_sH_s/2$, where $\mu_0$ is the vacuum permeability, $M_s$ the saturation magnetisation, and $H_s$ is the saturation magnetic field. We estimate from the hysteresis loop that $H_s$ in-plane is 0.8 T (based on when the magnetisation reaches 97.5\% of the fitted saturation value), and therefore $K_\text{eff} = 0.4\pm0.1$ MJ/m$^3$. The effective anisotropy includes the uniaxial and shape anisotropy.

For samples grown at intermediate temperatures with the chemically disordered $A1$ structure, the magnetism favours IP anisotropy at 40~nm, shown for growth at 550${\degree}$C in Fig. \ref{fig:Magnetization_Growth_T} (b).

For the chemically ordered $L1_0$ crystal structure shown in Fig. \ref{fig:Magnetization_Growth_T} (c), there is a significant increase in the coercivity and squareness ratio ($M_r/M_s$) for both the IP and OOP field orientations. The increased coercive field suggests that the $L1_0$ CoPt is magnetically hard compared to the $L1_1$ and $A1$ samples. The magnetisation of the $L1_0$ 40~nm CoPt sample does not show clear IP or OOP anisotropy from these measurements. 

The magnetisation vs growth temperature is shown in Fig. \ref{fig:Magnetization_Growth_T} (d). At growth temperatures below 550${\degree}$C the magnetisation remains approximately constant, however at higher temperatures the magnetisation begins to decrease with increasing temperature. The possible cause of this decrease is the higher growth temperature contributing to interdiffusion between the Pt and CoPt layers, creating magnetic dead layers.

The saturation field and squareness ratio vs growth temperature are shown in Fig. \ref{fig:Magnetization_Growth_T} (e) and (f) respectively. The general trends can be seen in the differences observed in the hysteresis loops of Fig. \ref{fig:Magnetization_Growth_T} (a-c). These trends allow us to characterise the growth temperatures corresponding to the different crystal growths in our samples, further supporting our conclusions from the XRD: the $L1_1$ crystal structure driving the magnetic response of samples grown from 200${\degree}$C to 400${\degree}$C, the $L1_0$ crystal structure for samples grown above 750${\degree}$C, and the intermediate temperatures being the chemically disordered $A1$ phase.

In the $L1_0$ crystal structure, Fig. \ref{fig:Magnetization_Growth_T} (c), the high squareness ratio for both field orientations suggests that the anisotropy axis of the material is neither parallel or perpendicular to the film. Instead, it is possible that the magnetic anisotropy is perpendicular to the layer planes, which are stacked in the [100] direction. 

\begin{figure}
    \centering
    \includegraphics[width=0.7\textwidth]{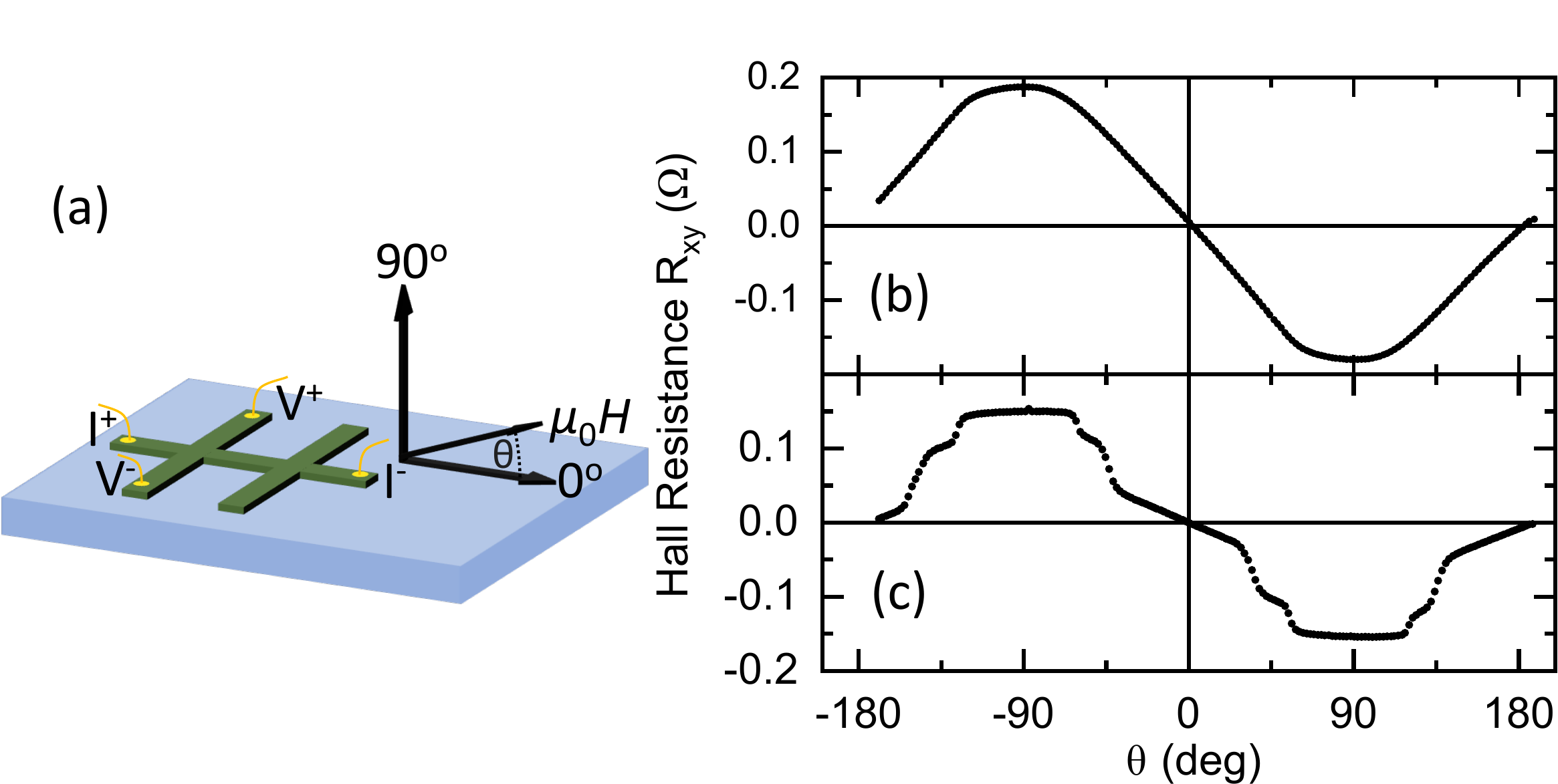}
    \caption{In-plane to out-of-plane angular dependent Hall resistance for Al$_2$O$_3$(sub)/Pt(4~nm)/CoPt(40~nm)/Pt(4~nm). (a) Schematic of Hall bar devices, measurement geometry, and applied field direction. Hall resistance verses angle for (b) the $L1_1$ crystal structure (350${\degree}$C growth temperature) at 0.4~T and (c) $L1_0$ crystal structure (800${\degree}$C) at 3~T. Data acquired at a temperature of 295~K.}
    \label{fig:Hallbar}
\end{figure}

To further investigate the anisotropy, we pattern our $L1_1$ 350${\degree}$C and $L1_0$ 800${\degree}$C samples into Hall bars and perform angular dependent Hall resistivity, R$_\text{xy}$($\theta$), measurements, Fig~\ref{fig:Hallbar}. The fabricated Hall bar and measurement geometry is shown in Fig~\ref{fig:Hallbar} (a). R$_\text{xy}$($\theta$) for the $L1_1$ 350${\degree}$C and $L1_0$ 800${\degree}$C Hall bars are shown in  Fig~\ref{fig:Hallbar} (b) and (c) respectively. For the $L1_1$ 350${\degree}$C sample with out-of-plane anisotropy, R$_\text{xy}$($\theta$) shows a plateau close to out-of-plane field and a uniform response for angles in between. The plateau is interpreted as an angle forming between the magnetisation and applied field because of the anisotropy axis \cite{doi:10.1063/1.3262635}. In comparison, the $L1_0$ 800${\degree}$C sample also shows a  R$_\text{xy}$($\theta$) plateau for out-of-plane applied field plus an additional plateau for applied field angles between about 45${\degree}$ and 60${\degree}$. We interpret the additional plateau in R$_\text{xy}$($\theta$) for the $L1_0$ 800${\degree}$C CoPt sample as evidence for an additional anisotropy axis, which we propose is perpendicular the [100] direction. The [100] plane has a dihedral angle of 54.75${\degree}$ with the [111] growth plane. Additional sources of anisotropy in our samples are interface anisotropy at the Pt/CoPt interfaces, which would favour out-of-plane magnetisation for thin layers, and shape anisotropy, which for our thin films would favour in-plane magnetisation.  

\begin{figure}
    \centering
    \includegraphics[width=0.5\textwidth]{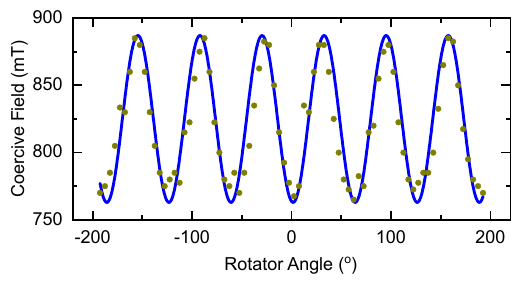}
    \caption{In-plane rotation of the Al$_2$O$_3$(sub)/Pt(4~nm)/CoPt(128~nm)/Pt(4~nm) sample with $L1_0$ crystal structure (800${\degree}$C). The coercive field is extracted from the hysteresis loop measured at each rotator angle. The position of 0$^\circ$ is arbitrary. The line shows a best fit sine function as a guide for the eye. Data acquired at a temperature of 300~K.}
    \label{fig:IP_Rotation}
\end{figure}

Fig. \ref{fig:IP_Rotation} shows the extracted coercive field as a function of in-plane rotator angle for the $L1_0$ 800${\degree}$C sample. The data shows a clear six-fold symmetry. This is consistent with an easy axis for the plurality tetragonal $L1_0$ phase of $[001]$ when grown on $\{111\}$ planes - the $<100>$ directions of the parent cubic structure are inclined at $\pm 45\degree$ from the plane and are coupled with the three fold symmetry of $\{111\}$. The magnetometry data therefore strongly suggests that over the sample the $[001]_{L1_0}$ can be found in any of the three possible $<100>$ of the parent cubic structure without a strong preference for which of the possible twins grow.


\subsection*{CoPt properties as a function of thickness for $L1_1$ and $L1_0$ crystal structures}

Having obtained optimal growth conditions for CoPt with chemically ordered $L1_1$ and $L1_0$ crystal structures, in this section, we report on the properties of samples Al$_2$O$_3$ (sub)/Pt (4~nm)/CoPt ($d_\text{CoPt}$)/Pt (4~nm) with $d_\text{CoPt}$ varied between 1~nm and 128~nm for both the $L1_1$ and $L1_0$ crystal structures.


\subsubsection*{\label{MagL11}Magnetisation of $L1_1$ and $L1_0$ CoPt}

\begin{figure}
    \centering
    \includegraphics[width=0.7\textwidth]{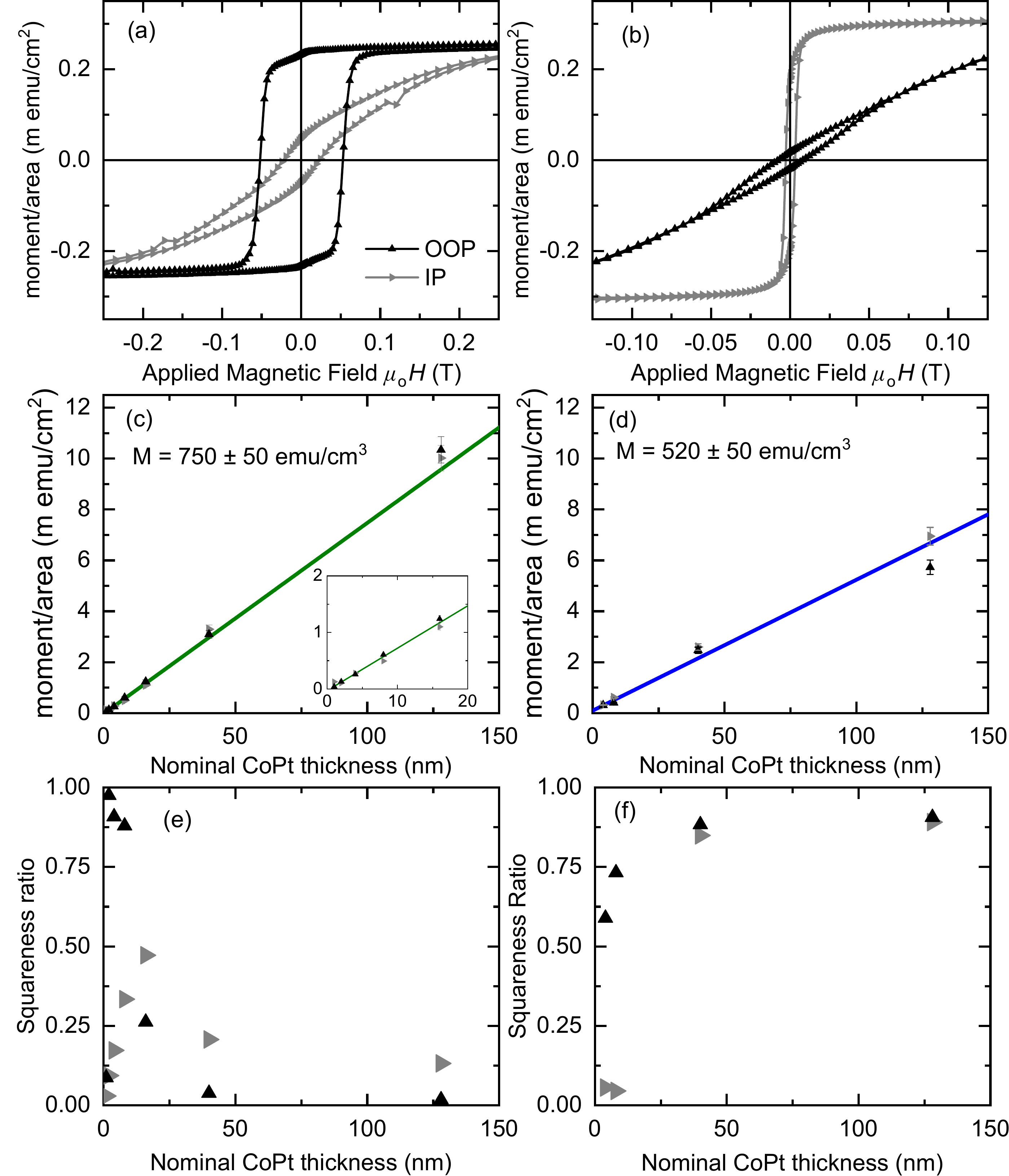}
    \caption{Magnetisation characterisation of Al$_2$O$_3$(sub)/Pt(4~nm)/CoPt($d_\text{CoPt}$)/Pt(4~nm) sheet films with the applied field oriented out-of-plane ($\blacktriangle$) and in-plane (\ulg). (a,b) Magnetic hysteresis loops acquired at a temperature of 300~K for $d_\text{CoPt}=4$~nm. The sample growth temperature is (a) 350${\degree}$C, corresponding to $L1_1$ crystal structure, (b) 800${\degree}$C, corresponding to $L1_0$ crystal structure. The diamagnetic contribution from the substrate has been subtracted. (c,d) Magnetic moment per area versus $d_\text{CoPt}$ for growth temperature of (c) 350${\degree}$C (with zoomed region inset) and (d) 800${\degree}$C. The significant contribution to the uncertainty in moment/area is from the area measurements of the sample portions. The magnetisation and uncertainty therein is determined from the fit to Eqn.~\ref{M2} (solid lines) and gives a best fit of $M=750\pm 50$~emu/cm$^3$ and $M=520\pm 50$~emu/cm$^3$ for the $L1_1$ and $L1_0$ crystal structures respectively. (e,f) The squareness ratio ($M_s/M_r$) versus $d_\text{CoPt}$ for growth temperature of (e) 350${\degree}$C and (f) 800${\degree}$C.}
    \label{fig:Magnetization_d}
\end{figure}

The hysteresis loops for Al$_2$O$_3$(sub)/Pt(4~nm)/CoPt(4~nm)/Pt(4~nm) sheet films is shown in Fig.~\ref{fig:Magnetization_d} (a) and (b) for the $L1_1$ (350${\degree}$C) and $L1_0$ (800${\degree}$C) crystal structures respectively. Hysteresis loops over the full thickness range are given in the Supplementary Information online. The moment/area at saturation (or 6~T) verses nominal CoPt thickness are presented in Fig~\ref{fig:Magnetization_d} (c) and (d).

We calculate the magnetisation ($M$) by fitting the moment/area versus nominal CoPt thickness data. In order to account for interfacial contributions to the magnetisation of the CoPt, we model the system as a magnetic slab with possible magnetic dead layers and/or polarised adjacent layers. Magnetic dead layers can form as a result of interdiffusion, oxidation, or at certain interfaces with non-ferromagnetic layers. At some ferromagnet/non-ferromagnet interfaces, the ferromagnetic layer can create a polarisation inside the non-ferromagnetic layer by the magnetic proximity effect. Polarisation is particularly common at interfaces with Pt \cite{doi:10.1063/1.344903, PhysRevB.65.020405, PhysRevB.72.054430,  rowan2017interfacial, PhysRevB.100.174418}. To take these into account, we fit to the expression,
\begin{equation}
\label{M2}
    \text{moment/area}= M  (d_\text{CoPt} - d_i),
\end{equation}
\noindent where $d_i$ is the contribution to $M$ from any magnetic dead layers or polarisation. The resulting best fit and the moment/area versus the nominal CoPt thickness is shown in Fig.~\ref{fig:Magnetization_d}.

For $L1_1$ growth at 350${\degree}$C, the result of fitting Eqn.~\ref{M2} shown in Fig.~\ref{fig:Magnetization_d} (c) gives $M=750\pm 50$~emu/cm$^3$ and $d_i=0.38\pm 0.05$~nm. From the XRR data and fitting presented in Fig.~\ref{fig:XRR}, the interdiffusion of the Pt seed and capping layers and the CoPt layer is minimal at this growth temperature, which is consistent with the small dead layer $d_i$.

For $L1_0$ growth at 800${\degree}$C, the thinnest 1~nm and 2~nm films did not display any magnetic response and are excluded from the analysis in Fig.~\ref{fig:Magnetization_d}. This suggests the formation of a magnetic dead layer or alternatively a large enough change to the stoichiometry that those films were no longer magnetic. From the XRR data and fitting presented in Fig.~\ref{fig:XRR}, there is significant interdiffusion between the Pt seed and capping layers and the CoPt. The result of fitting Eqn.~\ref{M2} shown in Fig.~\ref{fig:Magnetization_d} (d) for the samples thicker than 2~nm gives $M=520\pm 50$~emu/cm$^3$ and $d_i=-2\pm 1$~nm. Interestingly, we find a significant difference in $M$ between the two crystal structures, which is consistent with a previous report of CoPt growth on MgO substrates\cite{10.1063/1.4799526}. It is possible that the differences in $M$ corresponds to a true difference in the saturation magnetisation of the two crystal structures. An alternative scenario is that the 6~T applied field is not large enough to fully saturate the $L1_0$ samples, leading to a reduced measured $M$. Another possibility is that the interdiffusion of the Pt seed and capping layers during growth at 800${\degree}$C modifies the stoichiometry of the resulting $L1_0$ film, and hence reduces the magnetisation.

The thickness dependence of the magnetic switching of the $L1_1$ samples is well summarised by the squareness ratio shown in Fig~\ref{fig:Magnetization_d} (e). The thickest 40 and 128~nm samples are wasp waisted, as presented in Fig.~\ref{fig:Magnetization_Growth_T}. At reduced thicknesses, between 2 and 8~nm, the $L1_1$ CoPt no longer displays the wasp waisted switching for out-of-plane field orientation, and now has a square loop shown in Fig.~\ref{fig:Magnetization_d} (a). The 16~nm sample showed an intermediate behaviour. At 1~nm, the magnetic switching showed "S" shaped hysteresis loops for both in- and out-of-plane applied fields with small remanent magnetisation, see Supplementary Information online. 

The thickness dependence of the $L1_0$ crystal structure samples is significantly different to the $L1_1$. In the thinnest films of 1 and 2~nm there is no evidence of ferromagnetic ordering in the hystersis loops, see Supplementary Information online. For $L1_0$ growth at 800${\degree}$C, the XRR measurements (Fig.~\ref{fig:XRR}) suggests interdiffusion at the Pt/CoPt interfaces during high growth temperature. The interdiffusion may account for magnetic dead layers, which in the thinnest samples may prevent ferromagnetic ordering. Upon increasing the thickness to 4~nm, a ferromagnetic response was recovered, however the hysteresis loops and extracted squareness ratio (Fig.~\ref{fig:Magnetization_d}) (f)) indicate that the 4~nm and 8~nm $L1_0$ samples have in-plane magnetisation. The in-plane magnetisation in the thinner films suggests that the long-range $L1_0$ ordering may have not established at those thicknesses. 

The thickness dependence in both crystal structures suggest that the Pt/CoPt/Pt trilayers grown on Al$_2$O$_3$ substrates are not suitable for applications where ultrathin magnetic layers are required. To improve the magnetic properties of the thinnest samples in this study our future work will focus on replacing the Pt layers with seed and capping layers where interdiffusion may be less.


\subsection*{Spin polarisation}

To estimate the spin polarisation in the chemically ordered $L1_0$ and $L1_1$ CoPt samples, we perform Point Contact Andreev Reflection (PCAR) spectroscopy  experiments \cite{doi:10.1126/science.282.5386.85, baltz2009conductance, PhysRevB.85.064410, Seemann_2010, 6971749, PhysRevB.76.174435}. In the PCAR technique, spin polarisation in the ballistic transport regime can be determined from fitting the bias dependence of the conductance with the a modified Blonder-Tinkham-Klapwijk (BTK) model \cite{PhysRevB.63.104510}. 

\begin{figure}
    \centering
    \includegraphics[width=0.7\textwidth]{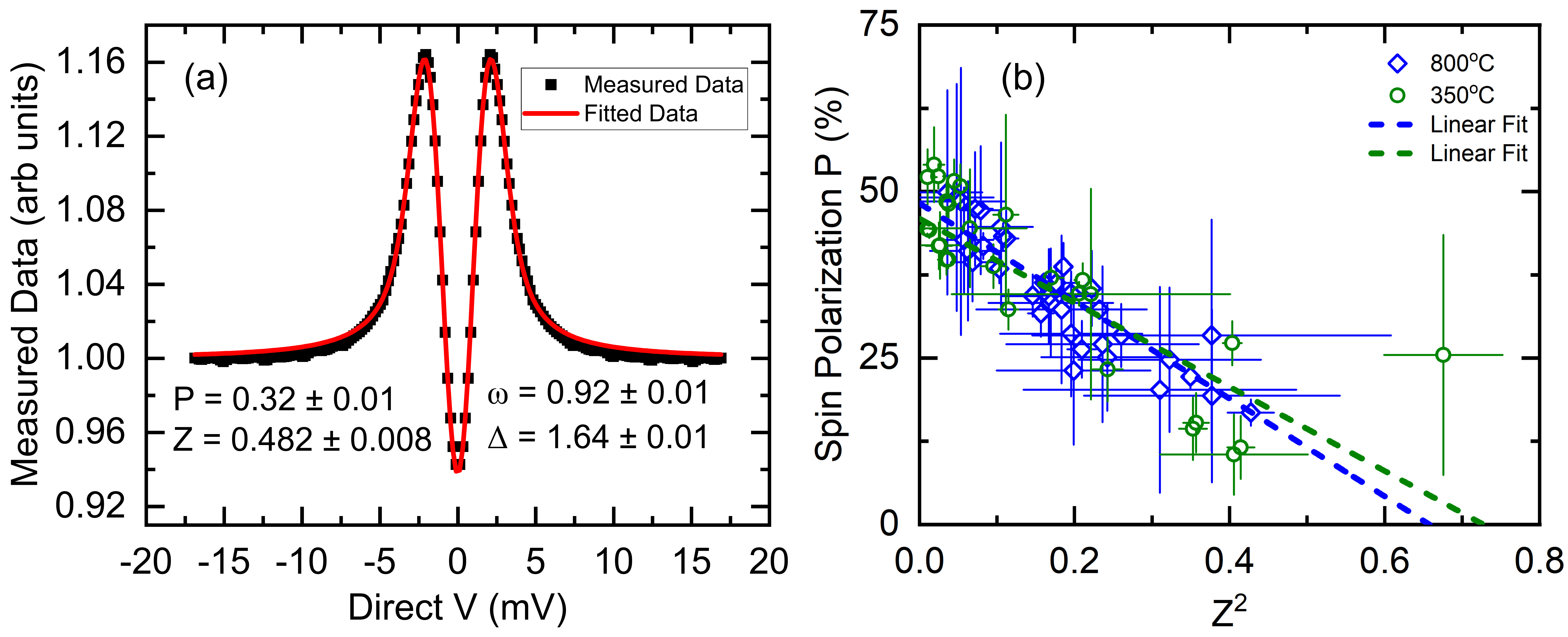}
    \caption{Point contact Andreev reflection measurements on the Al$_2$O$_3$(sub)/Pt(4~nm)/CoPt(128~nm)/Pt(4~nm) samples grown at 350${\degree}$C, corresponding to $L1_1$ crystal structure, and 800${\degree}$C, corresponding to $L1_0$ crystal structure. (a) Exemplar conductance versus voltage curve for the $L1_0$ sample with best fit to the BKT model (see text). (b) The polarisation is shown as a function of the square of the barrier strength, Z$^2$, with linear fits to the data.}
    \label{fig:PCAR}
\end{figure}

We measure the Al$_2$O$_3$(sub)/Pt(4~nm)/CoPt(128~nm)/Pt(4~nm) samples grown at 350${\degree}$C, corresponding to $L1_1$ crystal structure, and 800${\degree}$C, corresponding to $L1_0$ crystal structure. The PCAR experiment was performed with a Nb wire tip at $4.2\,$K. Exemplar conductance spectra with fits to the BTK model are given in Fig. \ref{fig:PCAR} (a). The interpretation of PCAR data is rife with difficulties\cite{Yates2018} and a common issue with the PCAR technique is the presence of degenerate local fitting minima. To ensure that a global best fit is obtained, the fitting code makes use of a differential-evolution algorithm and we then consider the spin polarisation and barrier strength parameter for a large number of independent contacts to the same sample.

Fig. \ref{fig:PCAR} (c) shows the dependence of the polarisation as a function of the square of the barrier strength, Z$^2$. The dashed lines in Fig. \ref{fig:PCAR} are linear fits to the data. The value of the true spin polarisation is is often taken to correspond to $Z=0$, however this is strictly nonphysical. Nevertheless, in an all metal system it is possible to produce contacts approaching an ideal case and extrapolating to $Z=0$ is close to the (finite) minimum. We find that P = 47$\pm3$\% for both $L1_1$ and $L1_0$ CoPt samples. This compares to $\approx42$\% for $L1_0$ FePt \cite{PhysRevB.76.174435} and $\approx50$\% for $L1_0$ FePd \cite{Seemann_2010}.


\section*{Conclusions}

The major conclusions of this work may be summarised as follows. On $c$-plane Al$_2$O$_3$ with thin Pt [111] buffer layers, the Co$_{50}$Pt$_{50}$ grows following the [111] ordering. Through growth at elevated temperatures, Co$_{50}$Pt$_{50}$ is grown epitaxially in the chemically ordered $L1_1$ and $L1_0$ crystal structures or the chemically disordered $A1$ phase. The $L1_1$ Co$_{50}$Pt$_{50}$ grown between 200$^\circ$C and 400$^\circ$C shows perpendicular magnetic anisotropy for thicknesses $\geq2$~nm. The $L1_0$ Co$_{50}$Pt$_{50}$ grown above 800$^\circ$C shows significantly harder magnetic anisotropy, which is perpendicular to the [100] direction for thicknesses $\geq40$~nm. For growth at intermediate temperatures, $450<800^\circ$C, the Co$_{50}$Pt$_{50}$ shows a disordered structure and in-plane magnetic anisotropy associated with the $A1$ phase. At growth temperatures of $550^\circ$C and above, significant interdiffusion between the CoPt and the Pt seed and capping layers is observed. The spin-polarisation of the $L1_1$ and $L1_0$ Co$_{50}$Pt$_{50}$ is determined by the PCAR technique to be 47$\pm3$\%.


\section*{Methods}


\subsection*{Epitaxial growth}

Samples are dc sputter deposited in the Royce Deposition System \cite{Royce}. The magnetrons are mounted below, and confocal to, the substrate with a source-substrate distance of 134~mm. The base pressure of the vacuum chamber is $1\times10^{-9}$~mbar. The samples are deposited at elevated temperatures, with an Ar (6N purity) gas pressure of $4.8\times10^{-3}$~mbar.

For alloy growth, we use the co-sputtering technique. To achieve as close to a Co$_{50}$Pt$_{50}$ stoichiometry as possible, first, single layer samples of Co or Pt are grown at room temperature on 10x10~mm thermally oxidised Si substrates varying the magnetron power. From this initial study, it is found that a growth rate of 0.05~nm s$^{-1}$ is achieved for a Co power of 45W and a Pt power of 25W. These growth powers are fixed for the rest of the study.

For the growth of the CoPt samples, 20x20~mm $c$-plane sapphire substrates are used. The substrates are heated by a ceramic substrate heater mounted directly above the substrate holder. The measured substrate heater temperature is reported. We note that the temperature on the substrate surface is most likely to be below the reported heater temperature. The substrate heater is ramped up from room temperature to the set temperature at a rate of 3-5${\degree}$C min$^{-1}$. Once at the set temperature, the system is given 30~min to reach equilibrium before starting the sample growth.

Once the system is ready for growth, 4~nm Pt seed layer is deposited. The seed layer is immediately followed by the  CoPt layer, which is deposited at a rate of 0.1~nm s$^{-1}$ by co-sputtering from the two targets at the determined powers. Finally, a 4~nm Pt capping layer is deposited to prevent the samples from oxidising. The final sample structure is Al$_2$O$_3$(sub)/Pt(4~nm)/CoPt($d_{CoPt}$)/Pt(4~nm). Following deposition, the samples are post growth annealed for 10~min at the growth temperature before the substrate heater is ramped down to room temperature at 10${\degree}$C min$^{-1}$.


\subsection*{Characterisation}

 Magnetisation loops are measured using a Quantum Design MPMS 3 magnetometer. Angular dependent magnetization measurements are performed using the Quantum Design Horizontal Rotator option. X-ray diffraction and reflectivity is performed on a Bruker D8 diffractometer with an additional four-bounce monochromator to isolate Cu K-$\alpha$ at a wavelength of 1.5406~\AA. Sheet films are patterned into Hall bars of 5~$\mu$m width using conventional photolithography and Ar ion milling. Resulting devices are measured using 4-point-probe transport to measure the Hall resistance of the films using a combined Keithley 6221-2182A current source and nano-voltmeter. 

\subsection*{PCAR}

Our experimental setup for performing PCAR measurements is described elsewhere \cite{baltz2009conductance, PhysRevB.85.064410, Seemann_2010, 6971749, PhysRevB.76.174435}. The Nb tips are prepared from commercial 99.9\% pure Nb wires with a diameter of 0.5~mm. An AC lock-in detection technique using Stanford Research Systems SR830 lock-in amplifiers is used for the differential conductance measurements. The tip position is mechanically adjusted by a spring-loaded rod driven by a micrometer screw. The experiment is carried out in liquid He at a fixed temperature of $4.2\,$K and at zero applied magnetic field.


\section*{Availability of Data and Materials}

The datasets generated and/or analysed during the current study are available in the University of Leeds repository, \url{https://doi.org/10.5518/1275}.


\bibliography{library}


\section*{Acknowledgements}

We wish to thank R. Hunt, M. Ali and  M. C. Rosamond for experimental assistance. We acknowledge support from the Henry Royce Institute. The work was supported financially through the following EPSRC grants: EP/V028138/1.


\section*{Author Contributions Statement}

N.S. and G.B. conceived and designed the study. N.S. with S.G., M.M., M.R., O.C. and G.B undertook the measurements and analysed the data. N.S. and P.M.S. undertook the materials development. N.S. undertook the sample growth and fabrication. N.S. with G.B. wrote the manuscript. All authors reviewed and edited the manuscript.


\section*{Additional Information}

\textbf{Supplementary information} accompanies this paper at URL.\\

\noindent \textbf{Competing interests:} The authors declare no competing interests.\\



\clearpage

\clearpage

\clearpage

\clearpage

\clearpage

\clearpage

\clearpage

\end{document}


\flushbottom
\maketitle
%
%
\thispagestyle{empty}

\section*{X-ray Diffraction Data}

\begin{figure}[ht]
\centering
\includegraphics[width=0.8\linewidth]{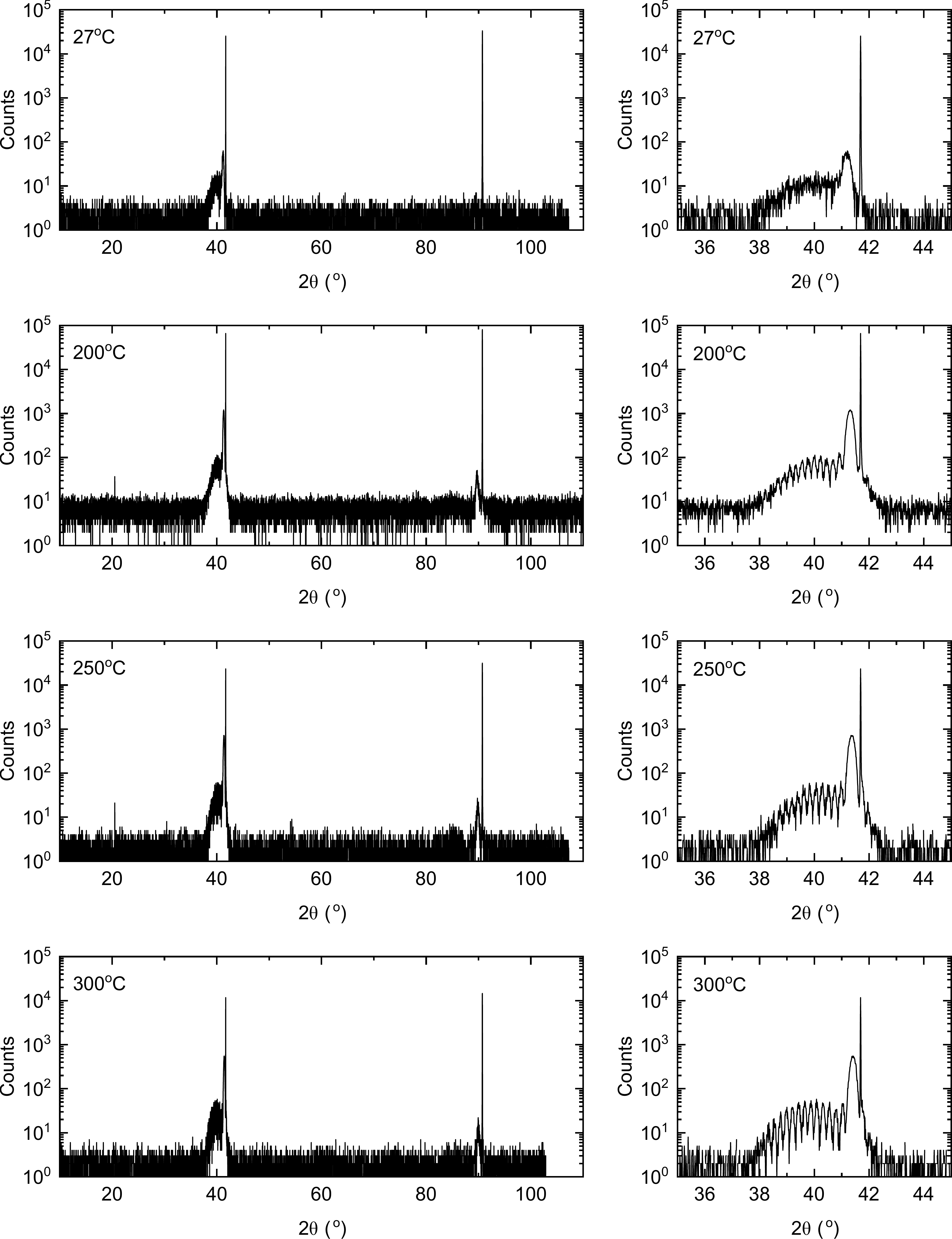}
\caption{X-ray diffraction for the growth temperature set of samples reported in the main text. Left column: Full data range. Right column: Cropped region around the 111 structural peak.}
\label{fig:XRD_ALL_1}
\end{figure}

\begin{figure}[ht]
\centering
\includegraphics[width=0.8\linewidth]{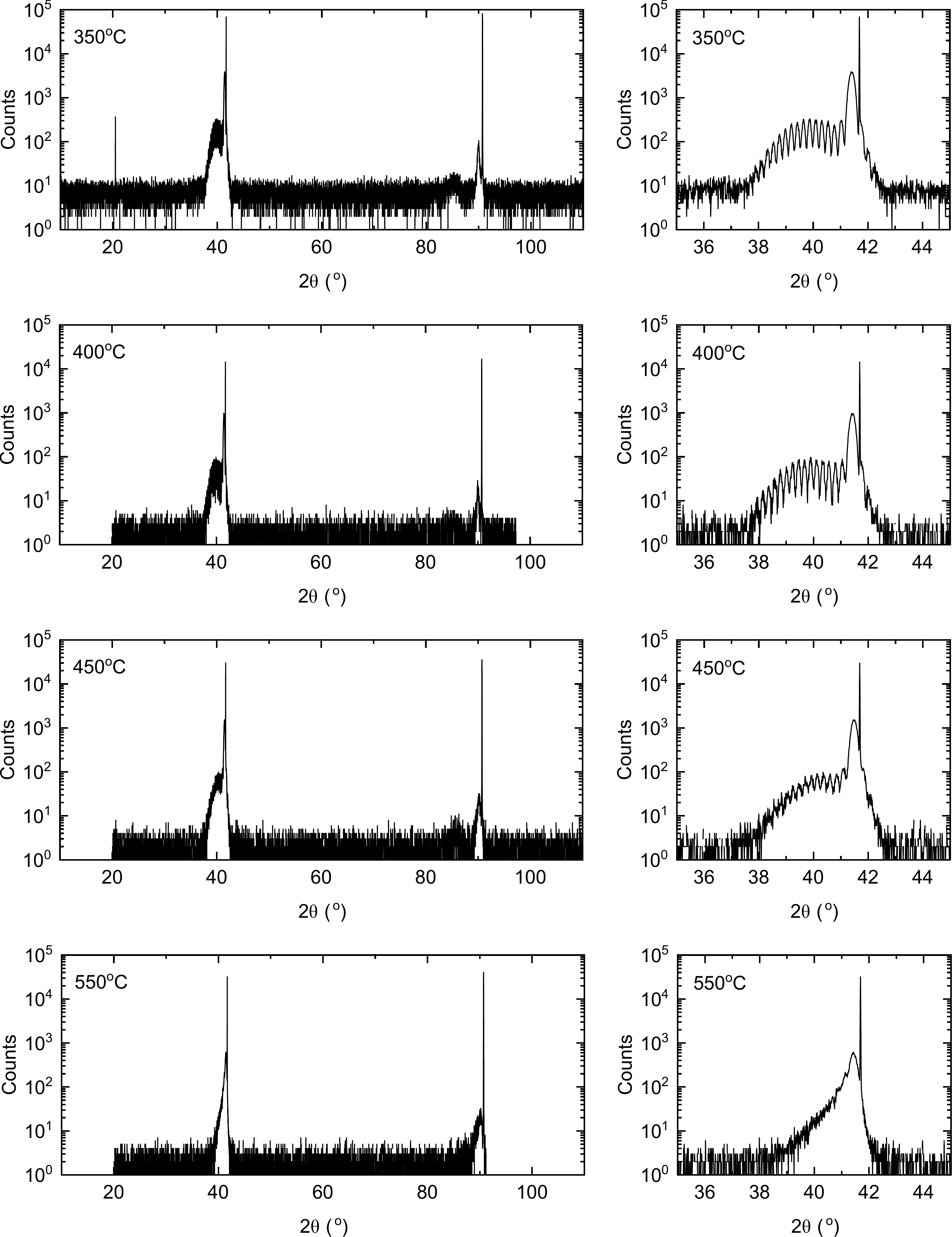}
\caption{X-ray diffraction for the growth temperature set of samples reported in the main text. Left column: Full data range. Right column: Cropped region around the 111 structural peak.}
\label{fig:XRD_ALL_2}
\end{figure}

\begin{figure}[ht]
\centering
\includegraphics[width=0.8\linewidth]{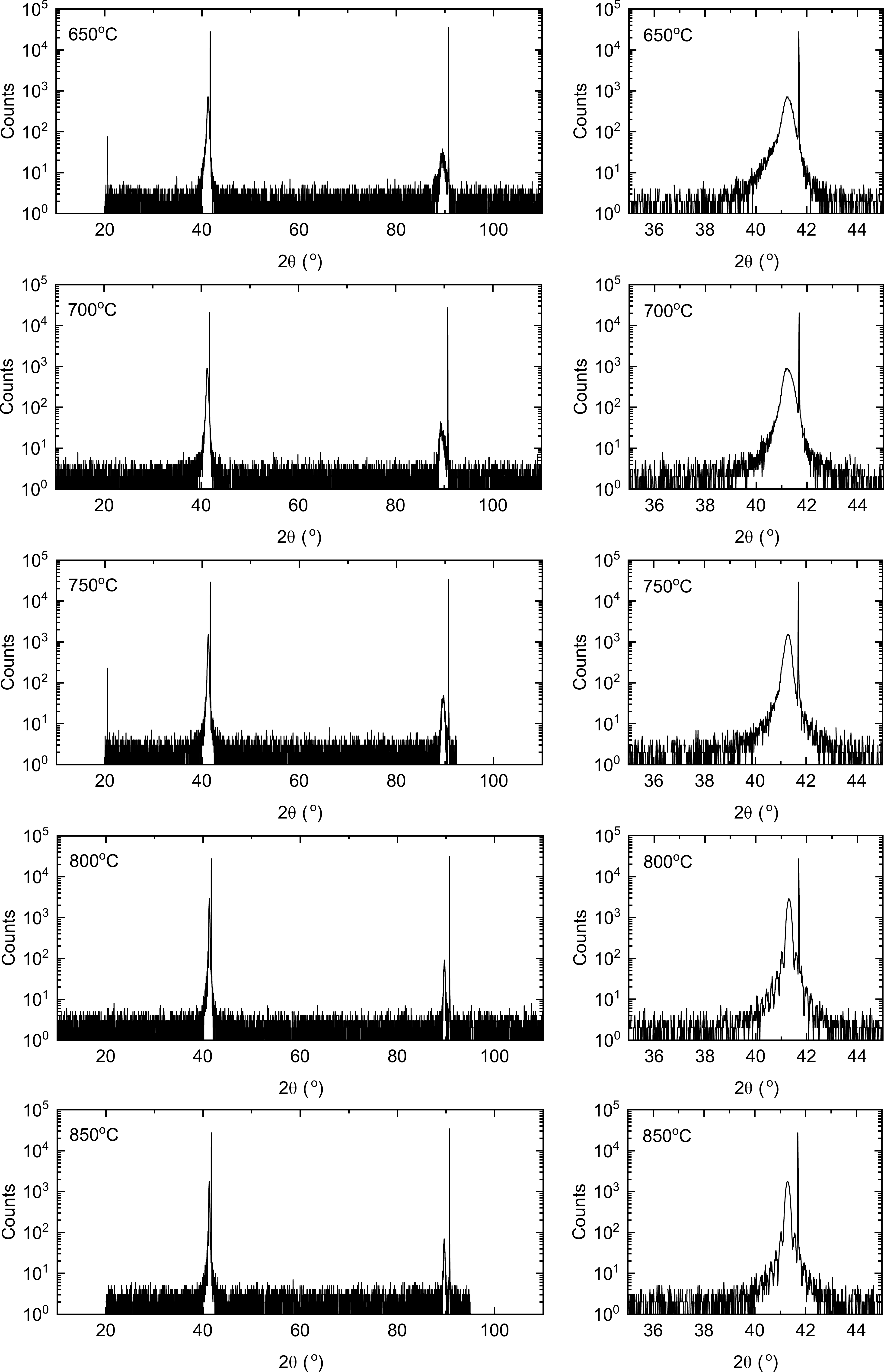}
\caption{X-ray diffraction for the growth temperature set of samples reported in the main text. Left column: Full data range. Right column: Cropped region around the 111 structural peak.}
\label{fig:XRD_ALL_3}
\end{figure}

\begin{figure}[ht]
\centering
\includegraphics[width=0.5\linewidth]{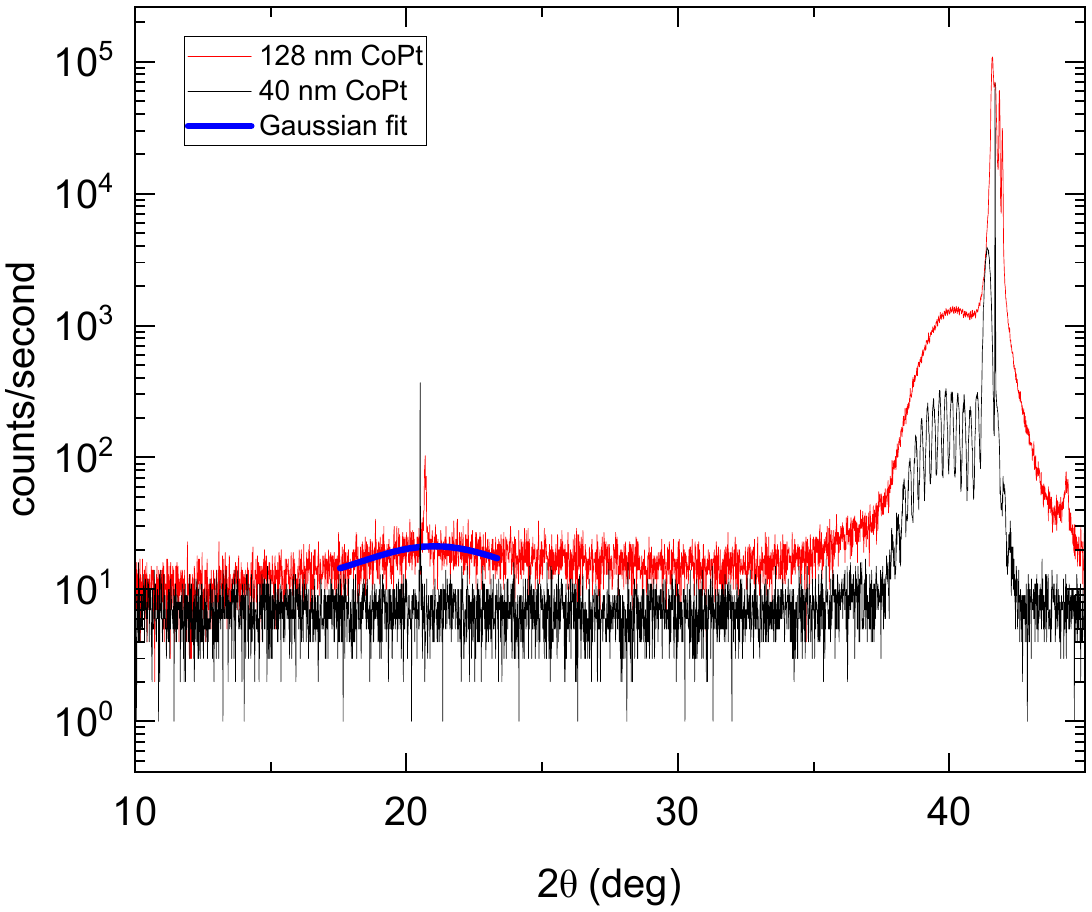}
\caption{X-ray diffraction for the 350$^\circ$C growth temperature over selected range. The 128~nm sample shows evidence of a superstructure peak at 2$\theta \approx 21^\circ$.}
\label{fig:350XRD}
\end{figure}

Figures \ref{fig:XRD_ALL_1}, \ref{fig:XRD_ALL_2}, \ref{fig:XRD_ALL_3} show a complete set of X-ray diffraction scans for samples grown at each temperature. The very sharp peak at $2\theta\approx42\degree$ is from the c-plane sapphire $(0\,0\,0\,6)$ plane with $\{1\,1\,1\}$ peaks from the CoPt phase clearly visible to left. The substrate $(0\,0\,0\,12)$ and CoPt $\{2\,2\,2\}$ peaks are also present close to $2\theta$ of $90\degree$. Samples grown at $200-450\degree\,$C and also at $800\degree\,$C clearly exhibit Pendellosung fringes, indicative of good crystal ordering from the L$1_1$ and L$1_0$ phases. The peak intensity and FWHM shown in the main text are extracted from a Gaussian peak fit to the CoPt $\{1\,1\,1\}$ peak data, although  similar data could be obtained from the CoPt $\{2\,2\,2\}$ data. Figure \ref{fig:350XRD} shows the XRD on the 40~nm and 128~nm samples grown at $350\degree\,$C. The additional peak at $\approx 21\degree$ in the 128~nm sample we interpret as the superstructure peak from the $L1_1$ growth.

\section*{X-ray Reflectivity Data}
\begin{table}[ht]
    \centering
    \begin{tabular}{|l|l|l|l|l|l|l|l|l|l|l|}
    \hline
        Growth Temp& Al$_{2}$O$_{3}$ & Pt1 & ~ & ~ & CoPt & ~ & ~ & Pt2 & ~ & \\ \hline
        ($\degree\,$C) & $\sigma$ (\AA)& \textit{d} (\AA)& $\sigma$ (\AA)& $\rho$ (f.u./\AA$^3$)& \textit{d} (\AA)& $\sigma$ (\AA)&$\rho$ (f.u./\AA$^3$)& \textit{d} (\AA)& $\sigma$ (\AA)& $\rho$ (f.u./\AA$^3$) \\ \hline
        27 & 4.9 & 42.8 & 5.6 & 0.0615 & 385 & 9.6 & 0.0758 & 40.9 & 8.0 & 0.0713  \\ \hline
        200 & 3.7 & 44.8 & 3.3 & 0.0615 & 382 & 10.8 & 0.0758 & 39.0 & 6.2 & 0.0713  \\ \hline
        250 & 3.3 & 46.0 & 3.9 & 0.0615 & 381 & 9.7 & 0.0758 & 40.8 & 5.8 & 0.0713  \\ \hline
        300 & 3.2 & 43.2 & 3.8 & 0.0615 & 388 & 13.5 & 0.0758 & 37.2 & 5.7 & 0.0713  \\ \hline
        350 & 2.8 & 42.1 & 4.1 & 0.0615 & 387 & 16.9 & 0.0758 & 36.4 & 4.8 & 0.0713  \\ \hline
        400 & 2.6 & 41.5 & 4.6 & 0.0615 & 388 & 15.7 & 0.0758 & 39.4 & 5.9 & 0.0713  \\ \hline
        450 & 3.3 & 41.1 & 4.1 & 0.0615 & 395 & 20.7 & 0.0758 & 32.2 & 7.5 & 0.0713  \\ \hline
        550 & 4.6 & ~ & ~ & ~ & 445 & 34.0 & 0.0758 & 29.6 & 8.9 & 0.0641  \\ \hline
        650 & 5.0 & ~ & ~ & ~ & 477 & 14.0 & 0.0812 & ~ & ~ &   \\ \hline
        700 & 5.4 & 7.6 & 4.2 & 0.0615 & 385 & 54.5 & 0.0758 & 61.6 & 19.1 & 0.0564  \\ \hline
        750 & 4.4 & ~ & ~ & ~ & 481 & 10.0 & 0.0786 & ~ & ~ &   \\ \hline
        800 & 5.4 & ~ & ~ & ~ & 479 & 9.5 & 0.0796 & ~ & ~ &   \\ \hline
        850 & 5.2 & ~ & ~ & ~ & 493 & 13.0 & 0.0724 & ~ & ~ &   \\ \hline
    \end{tabular}
\end{table}

\begin{figure}[ht]
\centering
\includegraphics[width=0.6\linewidth]{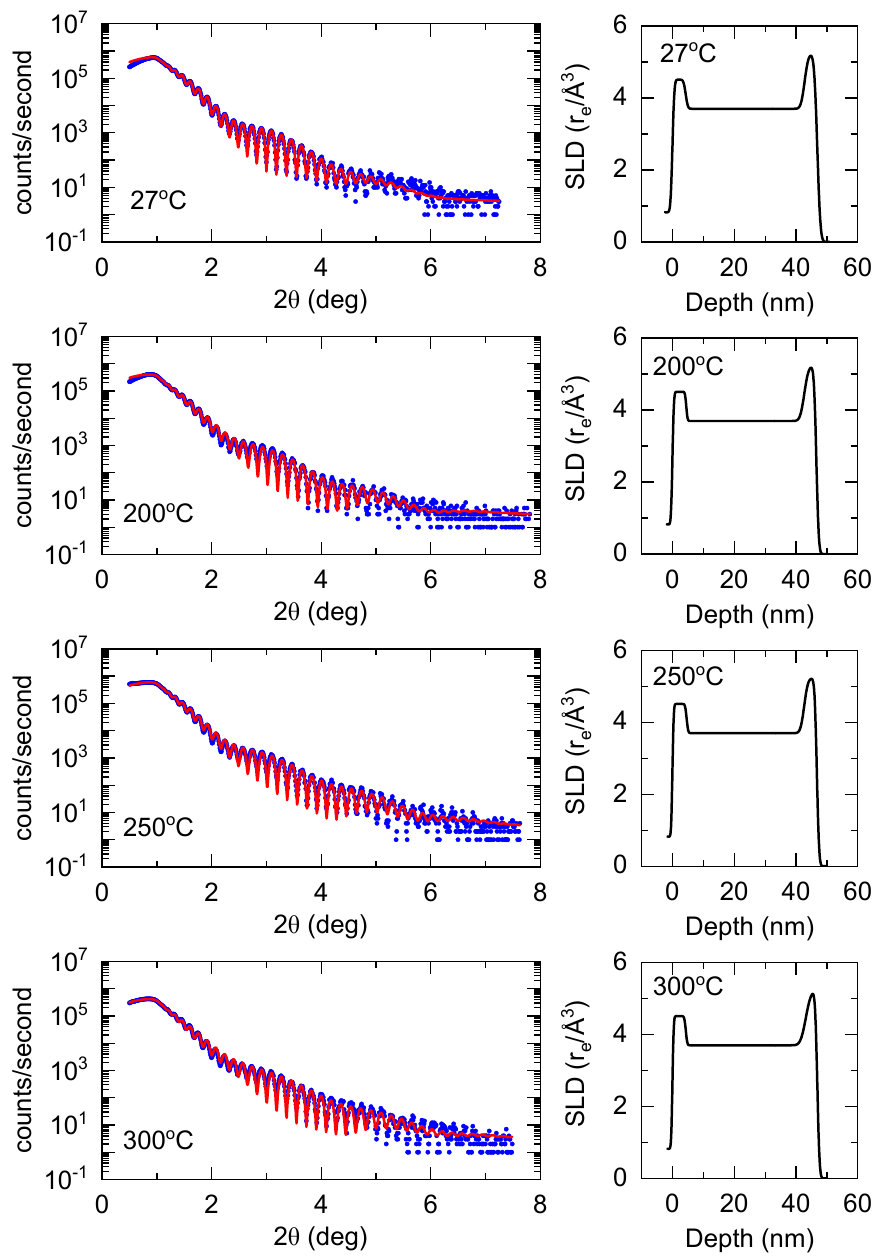}
\caption{X-ray reflectivity for the growth temperature set of samples reported in the main text. Left column: Full data range with best fit line. Right column: Structural model corresponding to best fit.}
\label{fig:XRR_ALL_1}
\end{figure}

\begin{figure}[ht]
\centering
\includegraphics[width=0.6\linewidth]{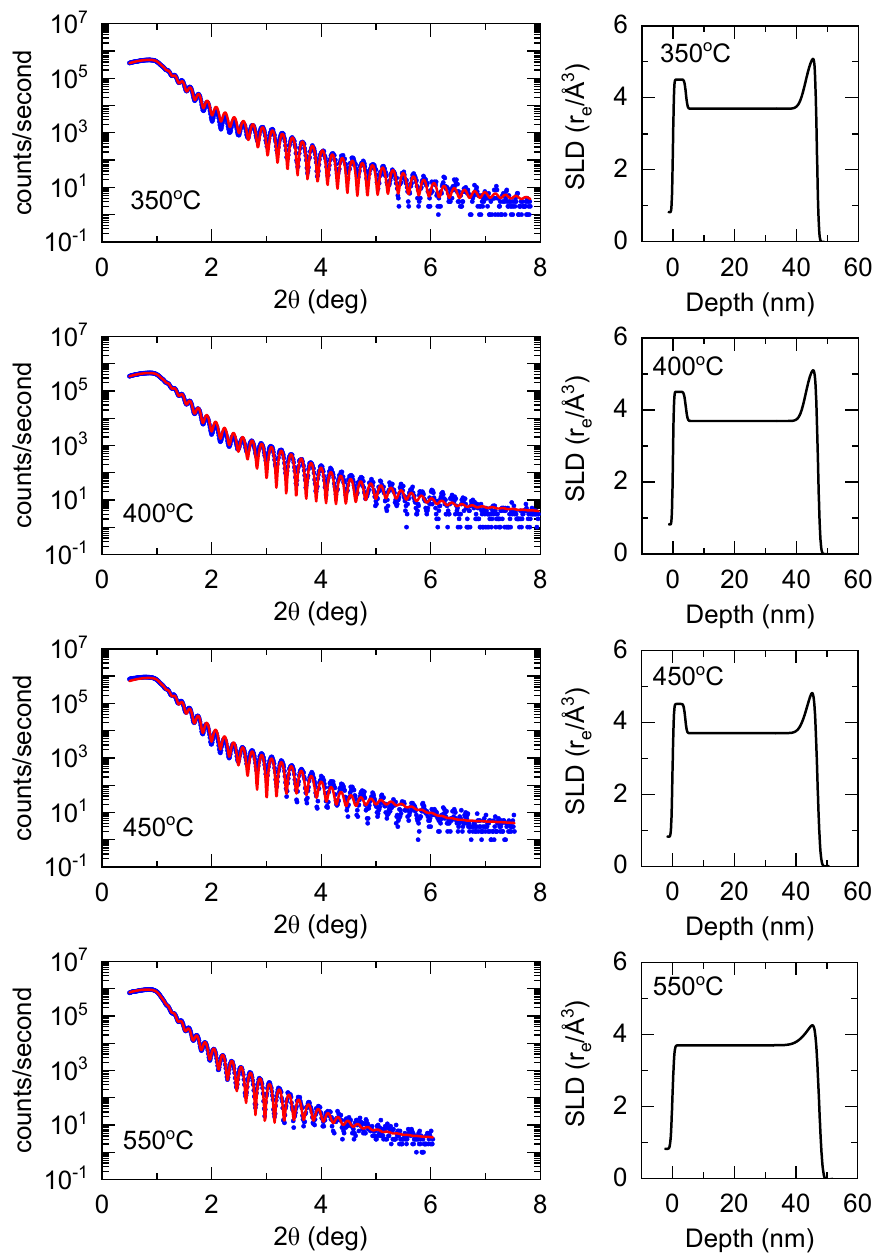}
\caption{X-ray reflectivity for the growth temperature set of samples reported in the main text. Left column: Full data range with best fit line. Right column: Structural model corresponding to best fit.}
\label{fig:XRR_ALL_2}
\end{figure}

\begin{figure}[ht]
\centering
\includegraphics[width=0.6\linewidth]{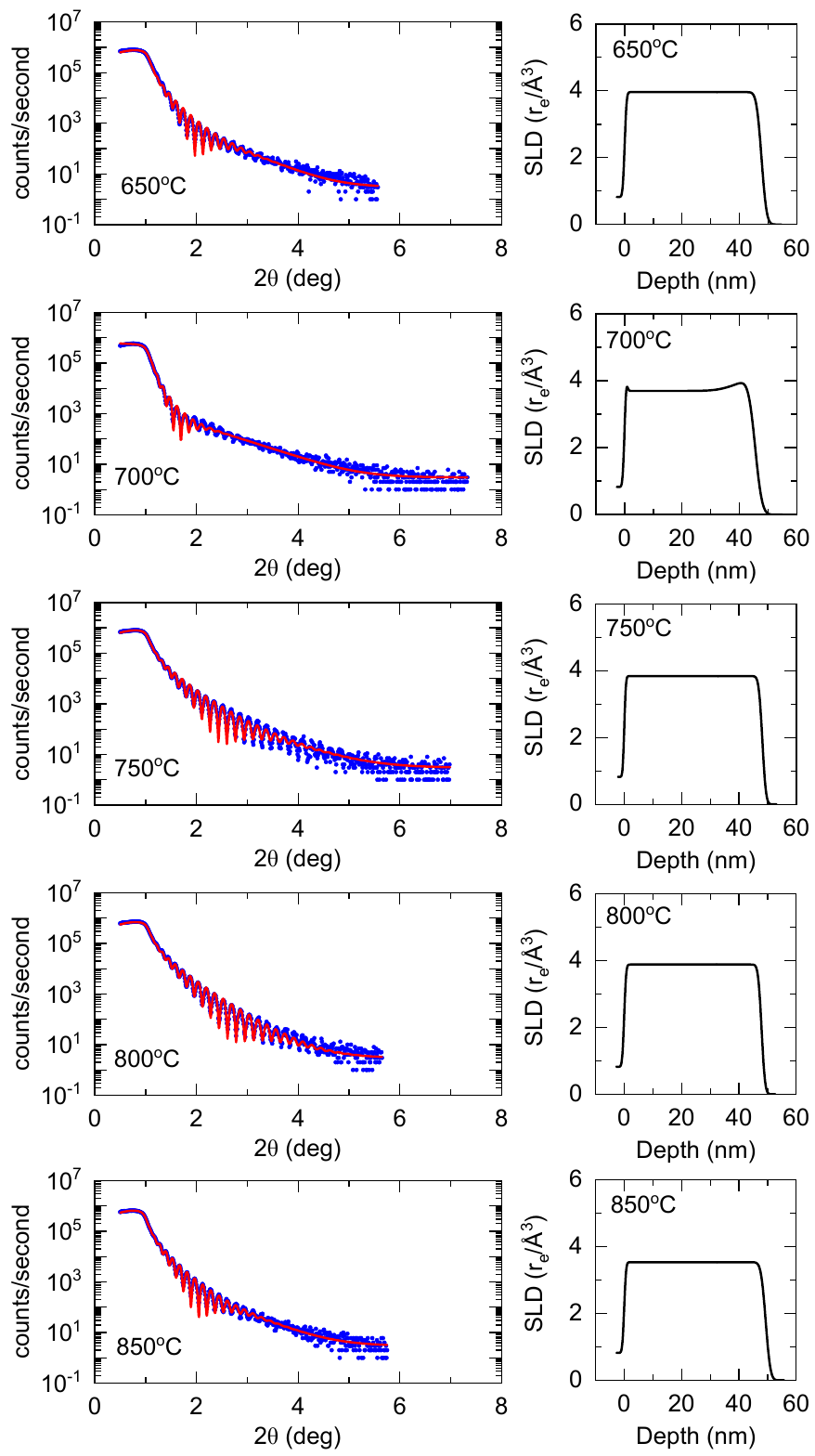}
\caption{X-ray reflectivity for the growth temperature set of samples reported in the main text. Left column: Full data range with best fit line. Right column: Structural model corresponding to best fit.}
\label{fig:XRR_ALL_3}
\end{figure}

Figures \ref{fig:XRR_ALL_1}, \ref{fig:XRR_ALL_2}, \ref{fig:XRR_ALL_3} show a complete set of X-ray reflectivity scans for samples grown at each temperature. The data with best fits are shown on the left, with the model corresponding to the best fit on the right side. The XRR are fit using the GenX package \cite{Bjorck:aj5091}. The extracted fitting parameters are given in the following table. Each layer is modelled as a box with a thickness ($d$) and a roughness ($\sigma$), in units of \AA, and a density ($\rho$), in units of formula unit (f.u.)/\AA$^3$. The substrate has only a roughness fitting parameter as the model treats it as in infinitely thick box with the bulk density of Al$_2$O$_3$. ``Pt1" layer is the seed layer and ``Pt2" is the capping layer. Between $27\degree\,$C and $450\degree\,$C, the best fit model is for the as-intended trilayer. Between these temperatures the density is fixed and only the thickness and roughness are free fit parameters. The slight difference in density between the seed and capping layer indicates that the top surface of ``Pt2" is oxidized. At higher temperatures the trilayer model breaks down and alternative fitting to either a single, bi-, or tri-layer model is used. 

\section*{Magnetometry}

\begin{figure}[ht]
\centering
\includegraphics[width=\linewidth]{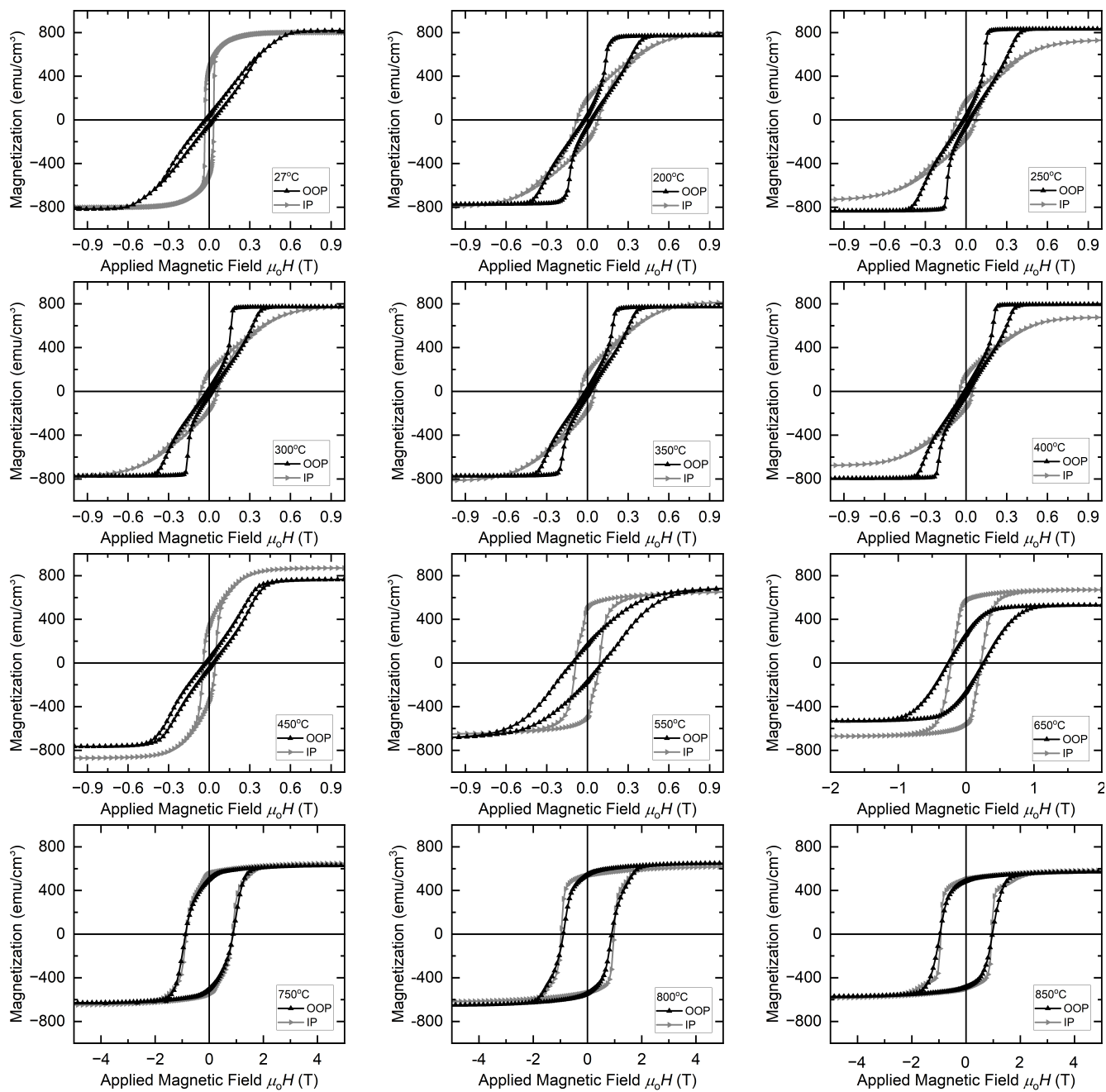}
\caption{Volume magnetisation for the growth temperature set of samples reported in the main text. The thickness of all samples is 40~nm.}
\label{fig:stream}
\end{figure}

\begin{figure}[ht]
\centering
\includegraphics[width=\linewidth]{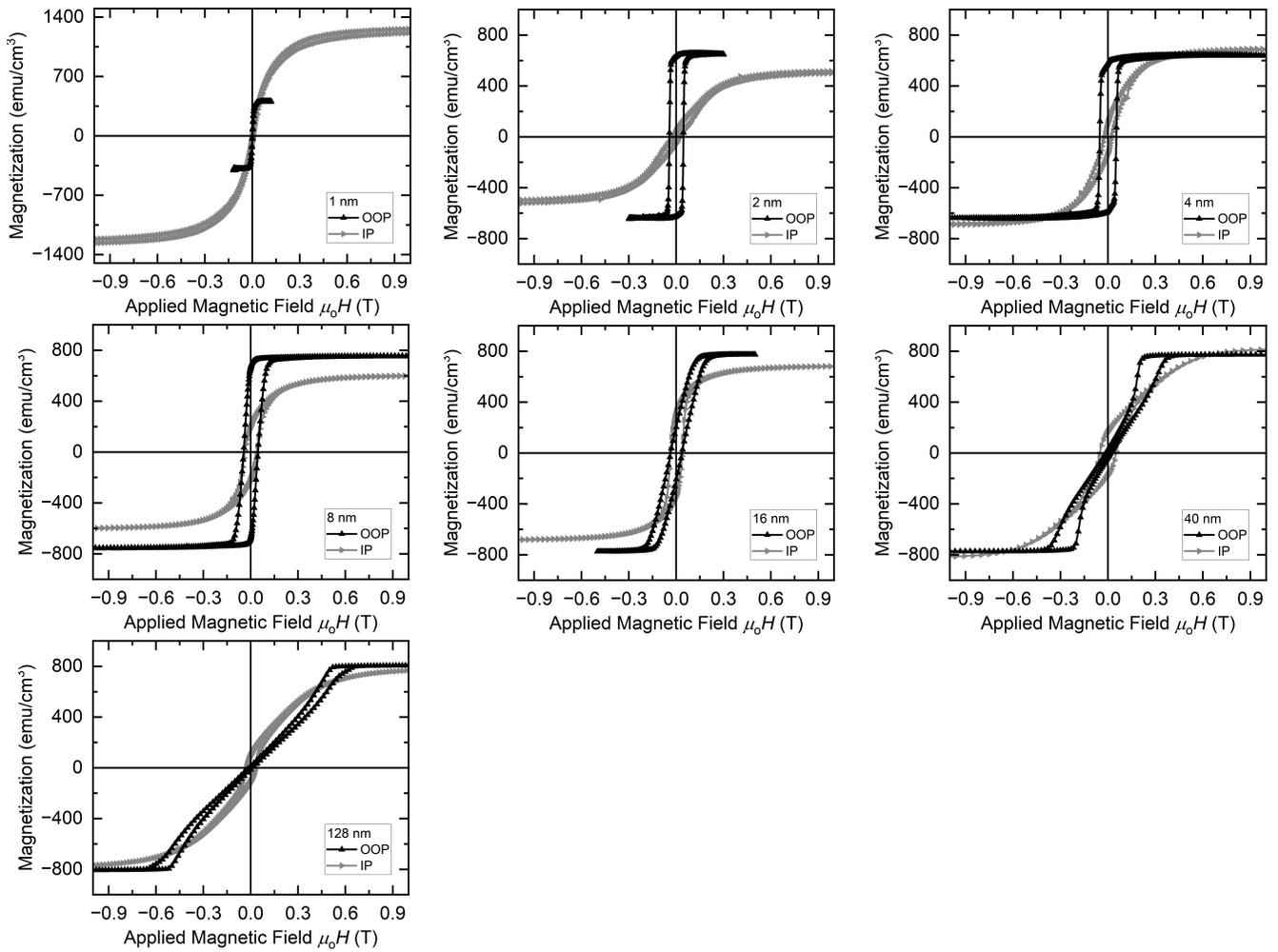}
\caption{Volume magnetisation for the thickness series set of samples reported in the main text for growth at 350$^\circ$C.}
\label{fig:stream2}
\end{figure}

\begin{figure}[ht]
\centering
\includegraphics[width=\linewidth]{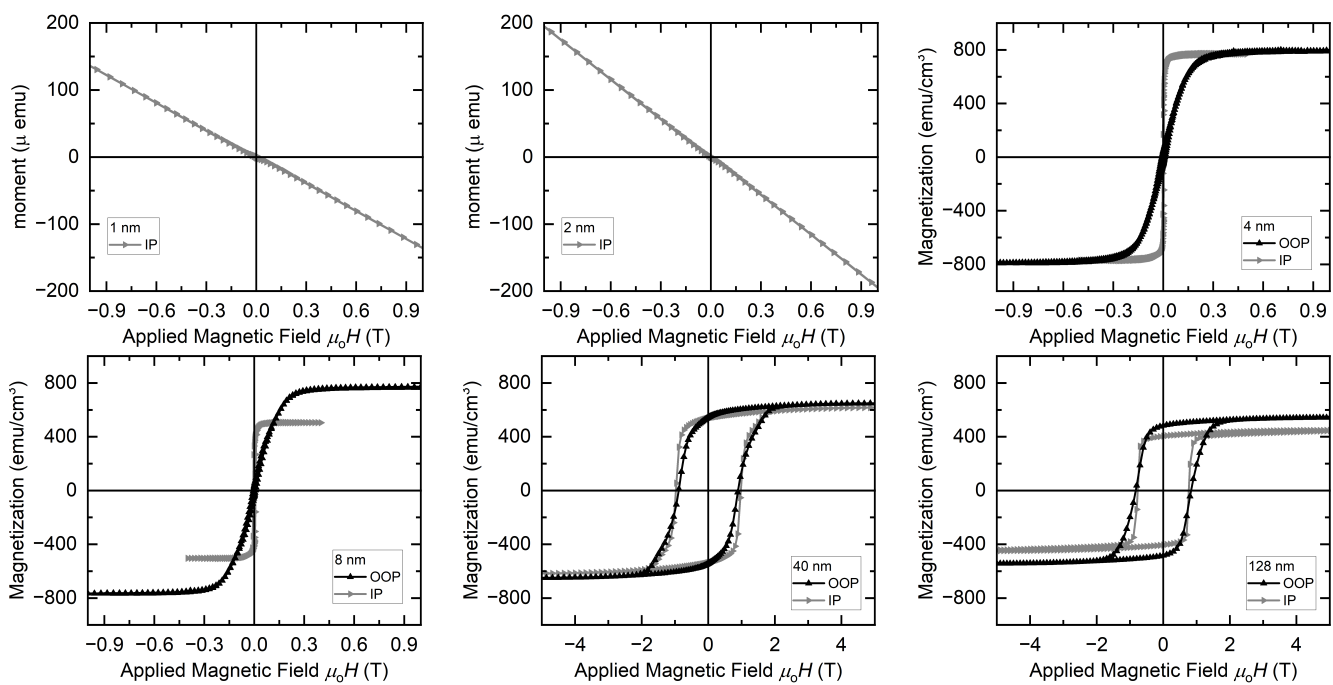}
\caption{Volume magnetisation for the thickness series set of samples reported in the main text for growth at 800$^\circ$C. For the 1~nm and 2~nm samples which did not show ferromagnetism, the uncorrected moment vs field is presented.}
\label{fig:stream3}
\end{figure}

Magnetic hysteresis loops were measured for a set of samples of common thickness grown at temperatures from room temperature to $850\degree\,$C with the field applied in the plane of the substrate and perpendicular to the substrate plane - shown in Figure \ref{fig:stream}. For the data shown here, a diamagnetic background resulting from the substrate and sample stick has been removed. The saturation magnetisation is determine by fitting a line to the outer section of the upper and lower branches and taking the difference in intercepts with the $H=0$ axis. The uncertainty is dominated by the determination of the sample area. The saturation field is determined from the point at which the measured data has deviated from the saturation magnetisation by more than twice the scatter in the data in the saturated regions of the loop. Figures \ref{fig:stream2} and \ref{fig:stream3} show similar magnetic hysteresis loops for samples of increasing thickness grown at $350\degree\,$C and $800\degree\,$C respectively.

\section*{PCAR}

The Point Contact Andreev Reflection spectra are measured by an AC lockin technique where the contact is subjected to a controlled DC bias with a small AC modulation which is sensed across the contact and across a control resistor to allow a differential conductance to be determined. The DC bias is separately monitored as it is adjusted to give a tip bias typically in the range of $\approx\pm30\,$meV. The DC bias is swept over 6 quarters from 0 to maximum and then to minimum and back to maximum and finally to zero. We only analyse the parts of the trace from maximum to minimum and back to maximum. The measured DC bias is subject to a small instrumentation offset which is removed at this first stage of data processing. Since all the features of interest in the spectra should be symmetrical about zero bias, we decompose the data into symmetric and anti-symmetric parts and retain only the symmetric part. The BTK model is usually considered in terms of the normalised conductance of the contact -- normalised to the conductance at high bias where transfer between electron/hole states in the ferromagnetic and electron/hole-like quasiparticles in the superconductor. In this high bias regime it is possible in some contact to get parabolic conductance due to tunnelling or Joule heating effects\cite{baltz2009conductance} and so we normalise the experimental data by dividing through a parabola fitted to the outer 20\% of the spectra.

The modified BTK model used in the analysis of the PCAR data has 4 fitting parameters. Firstly, an interfacial barrier strength $Z$ that is dimensionless. This effectively determines a scattering potential at the interface and so even in a perfect contact will take a small, but finite, value. Secondly, the spin polarisation $P=\frac{N_\uparrow v_{f\uparrow}-N_\downarrow v_{f\downarrow}}{N_\uparrow v_{f\uparrow}+N_\downarrow v_{f\downarrow}}$ which thus ranges from 0 for non-spin-polarised materials to 1 for fully spin-polarised materials. Thirdly, the superconducting energy gap $\Delta$ which might reasonably be expected not to be fully bulk like at the tip of the contact. Finally there is a `smearing parameter` $\omega$ that encompasses both thermal effects and athermal scattering processes not otherwise accounted for in the model.

The modified BTK model is then fitted to the now normalised data. We start with an initial set of fitting values of $Z=0.5$, $\omega=0.5\,$meV (c.f. $4.2\,$K $\equiv 0.36\,$meV), $\Delta=1.5\,$meV and $P=0.5$. We use a differential-evolution algorithm to identify a global minimum $\chi^2$ value to locate the best fit. This best-fit location is then used to seed a non-linar least squares fit using the Levenberg–Marquardt algorithm, allowing us to determine estimated standard errors for the fitting parameters.

We then discard fits with non-physical values of $\Delta$, excessibley large $\omega$ -- both of which would indicate an unaccounted spreading resistance, before considering all the fitted $(Z,P)$ combinations. As noted in the main text, the extrapolation back to $Z^2=0$ is not strictly physical, however, in practice in an all-metal system, very low values of $Z$ with uncertainties that encompass $Z=0$ can be obtained.

\bibliography{sample}